\documentclass[lettersize,journal]{IEEEtran}
\usepackage{amsmath,amsfonts}
\usepackage{algorithm}
\usepackage{array}
\usepackage[caption=false,font=normalsize,labelfont=sf,textfont=sf]{subfig}
\usepackage{textcomp}
\usepackage{stfloats}
\usepackage{url}
\usepackage{verbatim}
\usepackage{graphicx}
\usepackage{cite}
\usepackage{makecell}

\usepackage{amsthm,amsmath,amssymb}
\usepackage{makecell}
\usepackage{mathrsfs}
  \usepackage{float}
  \usepackage{algorithm}
\usepackage{algpseudocode}
\usepackage{amsmath}
\usepackage{bm}
\usepackage{color}
\usepackage{comment}
\usepackage{booktabs}
\usepackage{leftidx}
\usepackage{epstopdf}
\usepackage{amsmath}
\usepackage{makecell}
\usepackage{multirow}
\usepackage{hyperref}
\usepackage{amssymb}
\usepackage{adjustbox}
\begin{document}

\title{WiCo-PG: Wireless Channel Foundation Model for Pathloss Map Generation via Synesthesia of Machines}

\author{Mingran Sun,~\IEEEmembership{Graduate Student Member,~IEEE}, Lu Bai,~\IEEEmembership{Senior Member,~IEEE}, Ziwei Huang,~\IEEEmembership{Member,~IEEE}, Xuesong Cai,~\IEEEmembership{Senior Member,~IEEE}, Xiang Cheng,~\IEEEmembership{Fellow,~IEEE}, and Jianjun Wu,~\IEEEmembership{Member,~IEEE}

\thanks{M. Sun, Z. Huang, X. Cai, X. Cheng, and J. Wu are with the State Key Laboratory of Photonics and Communications, School of Electronics, Peking University, Beijing 100871, China (e-mail: mingransun@stu.pku.edu.cn; ziweihuang@pku.edu.cn; xuesong.cai@pku.edu.cn; xiangcheng@pku.edu.cn;
just@pku.edu.cn). 

L. Bai is with the Joint SDU-NTU Centre for Artificial Intelligence Research (C-FAIR), Shandong University, Jinan 250101, China (e-mail: lubai@sdu.edu.cn).}

}



\maketitle 

\begin{abstract}
A wireless channel foundation model for pathloss map generation (WiCo-PG) via Synesthesia of Machines (SoM) is developed for the first time. Considering sixth-generation (6G) uncrewed aerial vehicle (UAV)-to-ground (U2G) scenarios, a new multi-modal sensing-communication dataset is constructed for WiCo-PG pre-training, including multiple U2G scenarios, diverse flight altitudes, and diverse frequency bands. Based on the constructed dataset, the proposed WiCo-PG enables cross-modal pathloss map generation by leveraging RGB images from different scenarios and flight altitudes. In WiCo-PG, a novel network architecture designed for cross-modal pathloss map generation based on dual vector quantized generative adversarial networks (VQGANs) and Transformer is proposed. Furthermore, a novel frequency-guided shared-routed mixture of experts (S-R MoE) architecture is designed for cross-modal pathloss map generation. Simulation results demonstrate that the proposed WiCo-PG achieves improved pathloss map generation accuracy through pre-training with a normalized mean squared error (NMSE) of 0.012, outperforming the large language model (LLM)-based scheme, i.e., LLM4PG, and the conventional deep learning-based scheme by more than 6.98\,dB. The enhanced generality of the proposed WiCo-PG can further outperform the LLM4PG by at least 1.37\,dB using 2.7\% samples in few-shot generalization. 
\end{abstract}


\section{Introduction}
\IEEEPARstart{I}{n} wireless communication systems, channel modeling serves as a fundamental component of transceiver design and development \cite{channel modeling}. In general, channel fading characteristics can be categorized into large-scale and small-scale fading \cite{MMICM}. Specifically, large-scale fading primarily focuses on pathloss, which characterizes the spatial distribution of signal attenuation and serves as the foundation for coverage planning, power control, and interference management in system performance evaluation and network deployment \cite{large modeling, pathloss-twc, pathloss-twc2, pathloss-twc3}. Therefore, the investigation of pathloss is of paramount importance.

For conventional wireless communication systems from second-generation (2G) to fifth-generation (5G), pathloss generation approaches can generally be categorized into three types, named stochastic channel modeling approaches, ray-tracing (RT)-based approaches, and artificial intelligence (AI)-based radio-frequency (RF) data-driven approaches. The primary objective of these approaches is to provide fundamental support for link budgeting, coverage planning, and system-level simulations, while offering a unified benchmarking platform for algorithm performance evaluation. For conventional communication systems from 2G to 5G, pathloss generation accuracy requirements are relatively moderate, as long as the key channel statistical characteristics are captured. Specifically, stochastic channel modeling approaches \cite{scma1,  cai1, scma2, scma3,  cai3} utilized empirical formulas obtained by measurement data to generate pathloss data with low computational complexity. However, pathloss data obtained through stochastic channel modeling approaches have limited generation accuracy and cannot achieve high-quality pathloss generation. For the RT-based approaches, the electromagnetic wave propagation in detailed physical environments was mimicked to compute point-to-point pathloss \cite{RT1,RT2}. The RT-based approaches can offer high accuracy, whereas suffer from extremely high computational complexity, making massive data generation impractical. For the AI-based RF data-driven approaches \cite{RF-driven1, RF-driven2}, RF data was utilized to generate pathloss. Due to the lack of a detailed environmental understanding, the generation accuracy of AI-based RF data-driven approaches remains limited. Overall, the aforementioned three approaches in \cite{scma1, scma2, scma3, RT1,RT2,RF-driven1, RF-driven2} are sufficient for conventional communication system design and algorithm evaluation. However, with the advancement of communication system technologies, sixth-generation (6G) will deeply integrate AI into the core architecture design, forming an AI-native 6G communication system \cite{AI-native}. In such a condition, the performance ceiling of 6G AI-native communication systems fundamentally depends on the data quantity and quality \cite{AI-native}. Therefore, the demand for massive and high-quality pathloss data becomes critical. However, the stochastic channel modeling approaches and AI-based RF data-driven approaches cannot ensure high accuracy, while RT-based approaches cannot scale to massive pathloss generation due to their high computational cost. As a result, an urgent need exists for a novel pathloss generation framework capable of efficiently generating massive and high-quality pathloss data.

The aforementioned approaches in \cite{scma1, scma2, scma3, RT1,RT2,RF-driven1, RF-driven2} are intrinsically constrained by their exclusive dependence on RF information, without understanding the underlying physical environment. Therefore, the aforementioned approaches cannot simultaneously achieve massive and high-quality pathloss generation. To overcome this limitation, inspired by synesthesia of human, a novel concept, i.e., Synesthesia of Machines (SoM), was proposed in \cite{SoM}, which refers to the intelligent integration of communications and multi-modal sensing to enhance system performance. By exploiting the mapping mechanism between physical environment and electromagnetic space based on SoM, cross-modal pathloss generation can be achieved by easily accessible physical environmental data. In \cite{map1, map2, map3}, satellite maps were utilized to capture the global layout of buildings in a specific area. Furthermore, based on the convolutional neural networks (CNNs) and generative adversarial networks (GANs), pathloss maps were generated by exploring the mapping mechanism between environmental information in satellite maps and pathloss maps. However, the aforementioned work mentioned in \cite{map1, map2, map3} is limited to pathloss map generation in two-dimensional (2D) static ground-to-ground (G2G) communication scenarios. As 6G communication systems expand to dynamic and three-dimensional (3D) scenarios, such as uncrewed aerial vehicle (UAV)-to ground (U2G) scenarios, pathloss map generation becomes increasingly important as pathloss plays a crucial role in UAV networking and resource allocation \cite{U2G, uav-twc1, uav-twc2, uav-twc3}. However, the generation of pathloss maps becomes highly challenging due to 3D trajectories of UAVs and the high-mobility of physical scenarios. However, the global maps utilized in \cite{map1, map2, map3} only contain global layouts while lacking local environmental details of buildings that significantly affect electromagnetic propagation, making them unsuitable for pathloss generation in high-mobility 6G scenarios. To fill this gap, the sensing-based approach was proposed in our previous preliminary work \cite{mr_gan}, leveraging RGB-D images to capture fine-grained environmental details and generate pathloss maps in high-mobility U2G scenarios. Nevertheless, the aforementioned AI-based pathloss map generation work in \cite{RF-driven1, RF-driven2, map1, map2, map3, mr_gan} is fundamentally constrained by its dependence on conventional deep learning models. Due to architectural and parameter limitations, conventional deep learning models perform well under specific conditions whereas lack generalization capability across varying scenarios and frequency bands \cite{small-no}. Consequently, the accuracy and generalization of the conventional deep learning models degrade significantly in different scenarios, frequency bands, and flight altitudes, limiting their scalability and robustness for massive and high-quality pathloss data generation.

With the rapid development of natural language processing (NLP), large language models (LLMs) have demonstrated powerful capabilities in complex reasoning and multi-task generalization, outperforming conventional deep learning models \cite{LLM-yes, llm-twc}. LLMs have demonstrated strong capabilities in cross-modal generation tasks, including text-to-image and text-to-video \cite{llm-image, llm-video}, and have also inspired exploratory studies applying LLMs to channel-related tasks, such as channel estimation, prediction, and feedback \cite{llm4wm, llm-csi, llm-beam}. Through task-specific fine-tuning, LLMs can retain their powerful sequence modeling abilities while being adapted to specialized tasks. For pathloss map generation, an LLM-based pathloss map generation model (LLM4PG) was proposed in our previous work \cite{LLM4PG}. In \cite{LLM4PG}, RGB-D images were utilized to generate pathloss maps by fine-tuning a pre-trained LLM, i.e., GPT-2, for general knowledge transfer. 

Although LLM-based approaches exhibit superior reasoning and generalization capabilities compared with conventional deep learning models, they are constrained by the significant gap between the language domain and the multi-modal sensing-communication domain. As a result, the general knowledge transfer cannot be effectively specialized for cross-modal pathloss generation tasks, particularly when involving communication frequency–dependent electromagnetic characteristics. Given these constraints, foundation models pre-trained on extensive multi-modal data, i.e., sensing data and channel data, offer superior adaptability, unified cross-modal representation capability, and scalable deployment potential for cross-modal pathloss generation tasks \cite{channelgpt}. Therefore, to enhance the performance of 6G AI-native communication systems, building a wireless channel foundation model (WiCo) for cross-modal pathloss generation is essential for achieving massive and high-quality data generation under diverse conditions \cite{tnse}. However, compared with LLM-based approaches, the development of WiCo faces significant challenges in dataset construction, requiring a much larger amount of data. In addition, it demands a dedicated pre-training strategy for cross-modal pathloss generation, and network architectures better suited for generation and generalization across different scenarios, frequency bands, and flight altitudes.

To fill this gap, a novel WiCo for pathloss map generation (WiCo-PG) via SoM is proposed. The proposed WiCo-PG is pre-trained and evaluated in 3D high-mobility U2G scenarios. Based on the construction framework of the SynthSoM dataset \cite{synthsom}, a new multi-modal sensing-communication dataset is constructed for the mapping mechanism exploration between RGB images and pathloss maps. A novel architecture designed for WiCo-PG is proposed for cross-modal pathloss map generation. Furthermore, the proposed WiCo-PG serves as an effective tool for generating massive and high-quality pathloss maps, facilitating the development of 6G AI-native communication systems. The major contributions and novelties
of this paper are summarized as follows.

\begin{enumerate}
\item A novel WiCo-PG via SoM is developed, aiming to generate massive and high-quality pathloss maps required for 6G AI-native communication system design. Based on the designed network architecture and pre-training strategy, the proposed WiCo-PG can generate pathloss maps according to various RGB image and frequency inputs in diverse flight altitudes and frequency bands in multiple scenarios. Moreover, the propsed WiCo-PG achieves improved generation accuracy and enhanced generalization performance across diverse frequency bands and flight altitudes in multiple scenarios.

\item A new multi-modal sensing-communication dataset for WiCo-PG pre-training is constructed to explore the mapping mechanism between RGB images and pathloss maps. The constructed dataset contains 19.2\,k RGB images and 0.10\,B link-level pathloss values, covering two typical U2G scenarios, including urban crossroad and wide lane, each scenario with three frequency bands, including 1.6\,GHz, 15\,GHz, and 28\,GHz, and three flight altitudes, including 50\,m, 70\,m, and 80\,m.

\item A novel WiCo-PG network architecture is designed for cross-modal pathloss map generation, enabling the generation of pathloss maps from RGB images and frequency inputs across diverse conditions. Specifically, to realize cross-modal pathloss map generation and generalization across diverse conditions, a dual vector quantized generative adversarial networks (VQGANs) and Transformer-based architecture is proposed. Based on the proposed architecture, a frequency-guided shared-routed mixture of experts (S-R MoE) architecture is further designed. For the efficient pre-training of the proposed WiCo-PG, a two-stage pre-training strategy is introduced, which can establish collaborative representation and explore mapping mechanism between sensory synesthetic space and channel synesthetic space. 

\item Simulation results demonstrate that the proposed WiCo-PG achieves a normalized mean squared error (NMSE) of 0.021 in full-sample single condition pathloss generation, and achieve an NMSE of 0.012 in multi-condition dataset unified learning, outperforming the LLM4PG by more than 6.98\,dB. For the generalization performance, the proposed WiCo-PG shows promising zero-shot generalization performance across different scenarios, frequency bands, and flight altitudes. Furthermore, the proposed WiCo-PG can outperform the LLM4PG by at least 1.37\,dB using 2.7\% samples in few-shot generalization.

\end{enumerate}

The remainder of this paper is organized as follows. In Section II, the multi-modal sensing-communication dataset constructed for WiCo-PG pre-training is elaborated. Section III describes the network architecture of the proposed WiCo-PG designed for pathloss map generation. In Section IV, the simulation results are presented, where the full-sample pathloss map generation performance and the generalization performance across diverse conditions of the proposed WiCo-PG are analyzed and evaluated. Finally, conclusions are presented in Section V.

\section{Dataset Construction}
\begin{figure*}
\centerline{\includegraphics[width=0.95\textwidth]{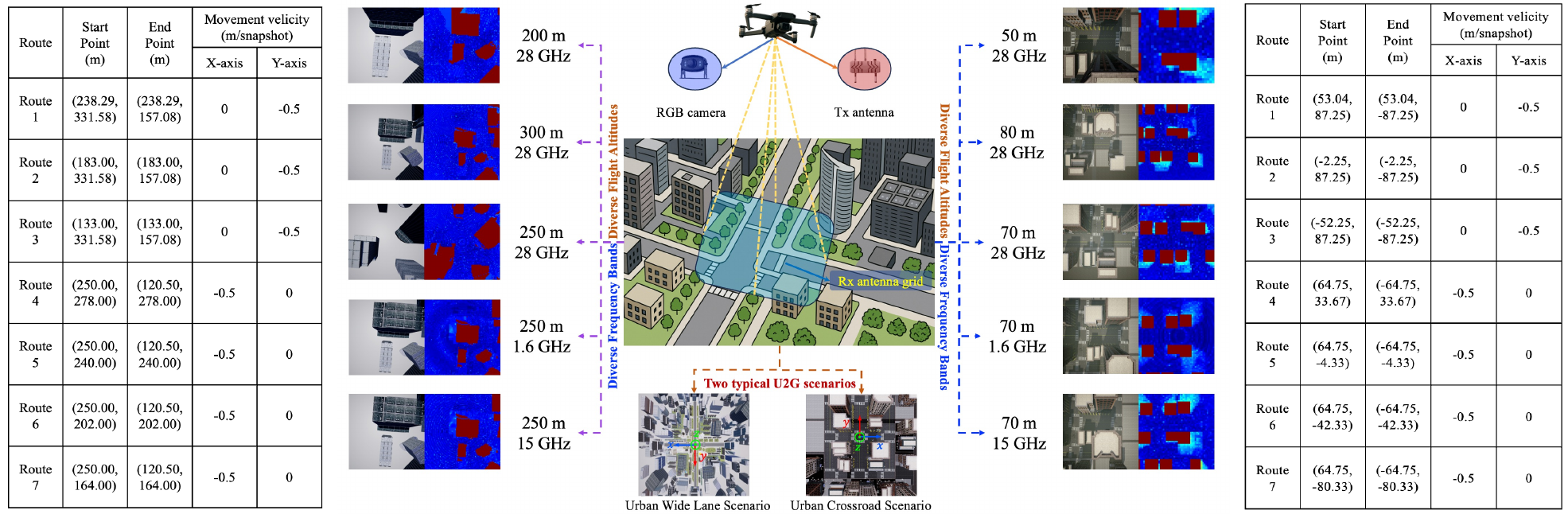}}
\caption{The constructed multi-modal sensing-communication dataset for WiCo-PG pre-training.\label{dataset}}
\end{figure*}

Based on the construction framework of the SynthSoM dataset \cite{synthsom}, a new multi-modal sensing-communication synthetic dataset is constructed for WiCo-PG pre-training. The dataset contains multi-modal data collected in U2G scenarios, including RGB images and pathloss data. To enable multi-condition dataset unified learning and cross-condition generalization testing for the WiCo-PG, the constructed dataset incorporates multi-modal data with diverse flight altitudes and frequency bands in multiple scenarios. Specifically, the constructed dataset comprises 19,150 pairs of RGB images and pathloss maps, covering two typical urban scenarios, i.e., urban crossroad and wide lane. In each scenaio, there are three flight altitudes, i.e., 50\,m, 70\,m, and 80\,m, and three communication frequency bands, i.e., 1.6\,GHz, 15\,GHz, and 28\,GHz. As illustrated in Fig.~\ref{dataset}, a UAV collects RGB images and corresponding U2G pathloss maps over a certain area. The proposed WiCo-PG learns the cross-modal mapping mechanism between RGB images and pathloss map, enabling massive and high-quality pathloss map generation from RGB image inputs. Fig.~\ref{dataset} illustrates the schematic of data collection scenarios and examples of multi-modal data, including RGB images and pathloss maps. The left part of Fig.~\ref{dataset} shows the UAV flight trajectory in the urban wide lane scenario, while the right part illustrates the UAV flight trajectory in the urban crossroad scenario. The details of the constructed multi-modal sensing-communication synthetic dataset for the WiCo-PG pre-training are presented in Table~\ref{dataset-table}. The dataset construction process and details of the dataset are elaborated below.

\begin{table}[!t]
\renewcommand\arraystretch{1.25}
\caption{Details of the Constructed Multi-Modal Sensing-Communication Synthetic Dataset for the Foundation Model Pre-training.\label{tab:table1}}
\centering
\begin{tabular}{|c|c|c|c|c|}
\hline
\textbf{\makecell[c]{Scenario}} & \textbf{\makecell[c]{Flight \\ altitude}} & \textbf{\makecell[c]{Frequency \\ band}} & \textbf{\makecell[c]{Number of \\ U2G links}} &\textbf{\makecell[c]{Number of \\ snapshots}} \\ 

\hline
\multicolumn{1}{|c|}{\multirow{5}{*}{\makecell[c]{Urban \\ crossroad}}} & \multicolumn{1}{c|}{\multirow{1}{*}{50\,m}} & \multicolumn{1}{c|}{\multirow{1}{*}{\makecell[c]{28\,GHz}}} & \multicolumn{1}{c|}{\multirow{1}{*}{\makecell[c]{1.65\,M}}} & \multicolumn{1}{c|}{\multirow{1}{*}{\makecell[c]{1,830}}} \\ 

    \cline{2-5}
    \multicolumn{1}{|c|}{}& \multicolumn{1}{c|}{\multirow{3}{*}{\makecell[c]{70\,m}}} & \multicolumn{1}{c|}{\multirow{1}{*}{\makecell[c]{1.6\,GHz}}} &  \multicolumn{1}{c|}{\multirow{1}{*}{\makecell[c]{4.58\,M}}} & {1,830} \\
    \cline{3-5} 
    {} & {} & 15\,GHz & 4.58\,M & {1,830} \\
    \cline{3-5}
    {} & {} & 28\,GHz & 4.58\,M & {1,830} \\
    \cline{2-5}
    {} & 80\,m & 28\,GHz & 6.59\,M & {1,830} \\
\hline
\multicolumn{1}{|c|}{\multirow{5}{*}{\makecell[c]{Urban \\ wide lane}}} & \multicolumn{1}{c|}{\multirow{1}{*}{200\,m}} & \multicolumn{1}{c|}{\multirow{1}{*}{\makecell[c]{28\,GHz}}} & \multicolumn{1}{c|}{\multirow{1}{*}{\makecell[c]{12.8\,M}}} & \multicolumn{1}{c|}{\multirow{1}{*}{\makecell[c]{2,000}}} \\ 

    \cline{2-5}
    \multicolumn{1}{|c|}{}& \multicolumn{1}{c|}{\multirow{3}{*}{\makecell[c]{250\,m}}} & \multicolumn{1}{c|}{\multirow{1}{*}{\makecell[c]{1.6\,GHz}}} &  \multicolumn{1}{c|}{\multirow{1}{*}{\makecell[c]{16.2\,M}}} & {2,000} \\
    \cline{3-5} 
    {} & {} & 15\,GHz & 16.2\,M & {2,000} \\
    \cline{3-5}
    {} & {} & 28\,GHz & 16.2\,M & {2,000} \\
    \cline{2-5}
    {} & 300\,m & 28\,GHz & 20.0\,M & 2,000 \\
    
\hline

\end{tabular}
\label{dataset-table}
\end{table}

\subsection{Initial Scenario Construction}

For multi-modal sensing-communication data acquisition, AirSim \cite{AirSim} is employed to collect RGB images, while Wireless InSite \cite{Wireless InSite} is utilized to collect  pathloss data. During the collection process, spatial and temporal alignment between the sensing and communication data is maintained to ensure precise correspondence across modalities. In the AirSim platform, two representative urban scenarios, i.e., the crossroad and the wide lane, are first constructed for data acquisition. Specifically, the crossroad scenario is built upon the high-fidelity 3D model Modular Building Set provided by PurePolygons, simulating a multi-directional traffic intersection environment. The wide lane scenario is designed based on the real geographic information of Chang’an Avenue in Beijing, reflecting the wide road structure and traffic characteristics of a typical urban arterial road. In the Wireless InSite platform, two 3D scenario object models are simplified and converted into compatible formats, ensuring that their geometries, dimensions, and spatial alignments remained consistent with the physical environments in AirSim. As a result, the precise alignment between the physical environment and the electromagnetic space is guaranteed.

For parameter configurations, in the crossroad scenario, the UAV flight altitudes are set to 50\,m, 70\,m, and 80\,m, while in the wide lane scenario, they are set to 200\,m, 250\,m, and 300\,m, enabling the collection of RGB images and pathloss maps of diverse flight altitudes. The same configurations is maintained across both simulation platforms in AirSim and Wireless InSite to ensure spatial and temporal alignment of multi-modal data. In Wireless InSite, all surfaces of buildings and roads are modeled with concrete material properties, representing the typical reflection and attenuation characteristics of urban environments. Furthermore, to collect pathloss data of diverse multiple frequency bands, three carrier frequencies, including 1.6\,GHz, 15\,GHz, and 28\,GHz, are selected for simulations in each scenario. 

Through the comprehensive scenario construction and parameter configuration process, multi-modal data of diverse flight altitudes and frequency bands in multiple scenarios are collected, providing a solid data foundation for subsequent foundation model pre-training and generalization evaluation. In total, the constructed dataset consists of 19.2\,k RGB images and 0.10\,B link-level pathloss values, which are utilized for training and testing the WiCo-PG.

\subsection{RGB Image Data Collection}

This subsection aims to collect RGB image data under different scenarios and flight altitudes for subsequent cross-modal pathloss map generation. The process begins with flight trajectory design and scenario generation in batches. Based on the initial scenarios constructed in Section II-A, the UAV flight trajectories are defined with the 3D coordinates point by point in each snapshot and kept consistent between AirSim and Wireless InSite. A scripted batch modification of configuration files in Wireless InSite is implemented to ensure the synchronized dynamic movement of the transmitter (Tx) antenna and the UAV, generating scenario samples across snapshots. The UAV flies along predefined trajectories at a constant speed, with the $x$–$y$ plane projection of the trajectories remaining identical across different flight altitudes. To enhance data diversity, the flight routes are designed to cover a broad area, enabling rich visual perspectives and spatial information.

Then, RGB image data acquisition is performed. At each snapshot position, an RGB camera deployed on the underside of the UAV captured downward-looking images of the scenario. The field of view (FoV) of the RGB camera is configured to match the coverage area of the ground receiver (Rx) antenna grid at each corresponding flight altitude, ensuring precise spatial alignment between RGB images and pathloss maps. The RGB images are automatically collected and stored in AirSim along the predefined flight trajectory. In total, 19,150 RGB images are collected. Specifically, in the urban crossroad scenario, 1,830 RGB images are obtained at each of the 50\,m, 70\,m, and 80\,m flight altitudes, while in the urban wide lane scenario, 2,000 RGB images are captured at each of the 200\,m, 250\,m, and 300\,m altitudes.

\subsection{Pathloss Data Collection and Pre-processing}

This subsection aims to collect pathloss maps aligned with the RGB images, and to process it into pathloss maps serving as ground-truth labels for subsequent cross-modal pathloss map generation. First, communication simulation configuration and data acquisition are conducted in Wireless InSite. By leveraging the UAV trajectories synchronized with AirSim, the Tx antenna mounted beneath the UAV and the ground Rx antenna grid are dynamically simulated. The Rx side consists of a high-density antenna grid, which can act either as a distributed multiple-input multiple-output (MIMO) array or as a set of potential Rx points, enabling extensive parallel acquisition of pathloss data. The density and coverage of the Rx grid varies across scenarios and flight altitudes. Specifically, in the urban crossroad scenario, the grids are set to 30×30, 50×50, and 60×60 for 50\,m, 70\,m, and 80\,m flight altitudes, respectively. In the urban wide lane scenario, the grids are 80×80, 90×90, and 100×100 for 200\,m, 250\,m, and 300\,m flight altitudes, respectively. Automated scripts are employed to execute simulation tasks across three carrier frequencies, including 1.6\,GHz, 15\,GHz, and 28\,GHz, to collect pathloss data under diverse communication conditions.

Then, data pre-processing are performed. The raw data exported from Wireless InSite contains point-wise pathloss values (in dB) between the UAV and each ground antenna of the Rx. To facilitate alignment with RGB image data and model training, the pathloss range [0, 255\,dB] is linearly mapped to pixel intensities [0, 255], thereby generating 2D pathloss maps. Each pathloss map is then paired with its corresponding RGB image at the same UAV position, achieving precise cross-modal alignment between the physical environment and the electromagnetic space. In total, 19,150 pathloss maps with 0.10\,B link-level pathloss values are collected. Specifically, in the urban crossroad scenario, 9,150 pathloss maps are obtained, while in the urban wide lane scenario, 10,000 pathloss maps are collected. Therefore, a comprehensive set of RGB image and pathloss map samples with diverse frequency bands and flight altitudes in multiple scenarios are obtained, providing solid data foundation for the pre-training and cross-modal generation of the proposed WiCo-PG.

\section{Framework of the Proposed WiCo-PG}

\begin{figure*}
\centerline{\includegraphics[width=0.9\textwidth]{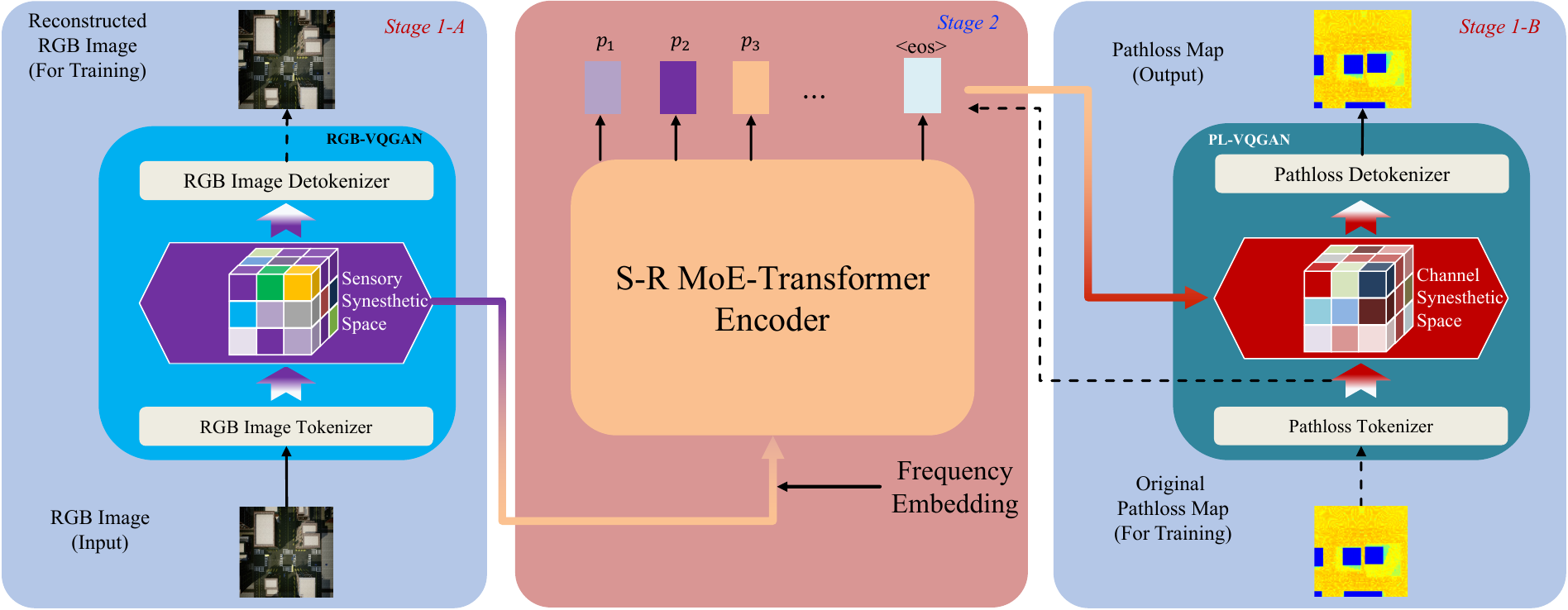}}
\caption{An illustration of the network architecture of the proposed WiCo-PG.\label{model}}
\end{figure*}

To achieve massive and high-quality cross-modal generation of pathloss maps from RGB images, based on the constructed dataset, a WiCo for pathloss map generation, named WiCo-PG, is proposed for the first time. However, the massive and high-quality cross-modal generation of pathloss maps faces several challenges, including the heterogeneous gap between RGB images and pathloss maps, strong domain shifts across different frequencies, flight altitudes, and scenarios, and the need for generalization under limited labeled data. To address these challenges, as shown in Fig.~\ref{model}, a dual-VQGAN and Transformer-based architecture composed of two stages and three modules is developed. In the stage 1, the RGB images and pathloss maps embedding modules project RGB images and pathloss maps into the sensory and channel synesthetic spaces, enabling unified feature representation. In the stage 2, the mapping mechanism exploration module explores the mapping mechanism between these two spaces. A frequency-guided S-R MoE Transformer enables cross-condition generalization. The detailed architecture of the proposed WiCo-PG will be elaborated below.

\subsection{Stage 1: Synesthetic Space Embedding}

\subsubsection{RGB Image Embedding Module}

To achieve cross-modal generation from RGB images to pathloss maps, the designed WiCo-PG first employs the dual-VQGAN network to map RGB images and pathloss maps into the sensory synesthetic space and the channel synesthetic space, respectively. Then, a Transformer is utilized to explore the intrinsic mapping mechanism between these two spaces. Specifically, the synesthetic space is defined as a discrete and shared representation space constructed within the cross-modal data generation framework, where sensory and communication modalities are represented through a unified encoding process. It can be divided into the sensory synesthetic space and the channel synesthetic space. To map RGB images into the sensory synesthetic space, an RGB image embedding module is designed and implemented to process RGB images, as shown in Stage 1-A of Fig.~\ref{model}. The module maps raw RGB images into the sensory synesthetic space, providing essential representations for subsequent cross-modal generation and mapping tasks.

The input image is a three-channel RGB image $\mathit{I} \in \mathbf{R}^{H \times W \times 3}$, containing the geometric structures and semantic layouts of the scenario. Since raw RGB images are high-dimensional, directly feeding them into the cross-modal generation model may lead to excessive computational cost and hinder convergence. To address this issue, a ResNet-based VQGAN \cite{vqgan} is employed to discretize the image features, achieving compact yet high-fidelity representations.
Specifically, the input RGB image is first encoded by an encoder $E_{\mathrm{V}}$ composed of multiple convolutional and residual blocks to extract hierarchical spatial features, yielding an intermediate continuous latent representation
\begin{equation}
\hat{z}^{S} = E_{\mathrm{V}}(I)\in \mathbf{R}^{h \times w \times n_z}
\end{equation}
where $h \times w \times n_z$ denotes the dimensionality of the continuous latent representation vector. A discrete codebook $\mathcal{Z}=\{{z_k}\}_{k=1}^{K}, z_k \in \mathbf{R}^{n_z}$ is pre-defined, in which each latent feature vector in $\hat{z}^{S}$ is replaced by its nearest codeword $z_q^{S}$ from $\mathcal{Z}$ through a quantization operation $\mathrm{Q}$, which is presented as 
\begin{equation}
z_q^{S} = \mathrm{Q}(\hat{z}^{S}, \mathcal{Z}).
\end{equation}
This process transforms the continuous latent features into a discrete latent representation $z_q^{S}$ of the same dimension, where both $\hat{z}^{S}$ and $z_q^{S}$ share the feature dimension $n_z$.
The discrete latent variable $z_q^{S}$ serves as the encoded representation of the RGB image in the sensory synesthetic space.
During training, the VQGAN is optimized by the loss function  $\mathcal{L}$ combining the vector quantization loss $\mathcal{L}_{\mathrm{VQ}}$ and the adversarial loss $\mathcal{L}_{\mathrm{GAN}}$, which is presented as
\begin{equation}
\mathcal{L} = \mathcal{L}_{\mathrm{VQ}} + \lambda \mathcal{L}_{\mathrm{GAN}}
\end{equation}
where
\begin{equation}
\begin{aligned}
\mathcal{L}_{\mathrm{VQ}} = & \; || I - \hat{I} ||^2 
+ || \mathrm{sg}[E_{\mathrm{V}}(I)] - z_q^{S} ||^2 \\
& + || \mathrm{sg}(z_q^{S}) - E_{\mathrm{V}}(I) ||^2
\end{aligned}
\end{equation}
and 
\begin{equation}
\begin{aligned}
\mathcal{L}_{\mathrm{GAN}} 
= & \; \log D(I)  + \log \big( 1 - D(\hat{I}) \big).
\end{aligned}
\end{equation}
Specifically, $\hat{I}$ denotes the reconstructed image, $\mathrm{sg}[\cdot]$ is the stop-gradient operation that ensures proper gradient flow through the encoder during backpropagation while preventing interference from the discrete quantization step, $\lambda$ is a weighting coefficient for the adversarial term, and $D$ denotes the discriminator network, which is trained to distinguish between the real RGB image $I$ and the reconstructed one $\hat{I}$. The discriminator in the adversarial loss adopts a hinge loss formulation to enhance reconstruction fidelity. Through this discrete encoding mechanism, RGB images are transformed into latent codes in the sensory synesthetic space, providing a representation for the subsequent cross-modal mapping process.



\subsubsection{Pathloss Map Embedding Module}

Pathloss represents the attenuation of electromagnetic signal power as it propagates from the Tx to the Rx through the environment \cite{pl}. It quantifies the combined effects of free-space spreading, reflection, diffraction, scattering, and absorption caused by environmental objects, such as buildings and terrain. Mathematically, pathloss is defined as the ratio between the transmitted power $P_{\mathrm{t}}$ and the received power $P_{\mathrm{r}}(d)$ at distance $d$, expressed in dB as
\begin{equation}
\mathrm{PL}(d) = 10 \log_{10} \left( \frac{P_{\mathrm{t}}}{P_{\mathrm{r}}(d)} \right).
\end{equation}
Pathloss is a key metric in wireless communications, directly influencing link budget estimation, coverage prediction, and resource allocation. A pathloss map spatially represents these attenuation values over a given area and is strongly influenced by the transceiver distance and the communication frequency band.

To encode pathloss maps into the channel synesthetic space, a pathloss embedding module is designed to process and discretize the spatial distribution of signal attenuation, as shown in Stage 1-B of Fig.~\ref{model}. Unlike RGB images that capture visual semantics, pathloss maps encode electromagnetic propagation characteristics that depend on geometry, material composition, and frequency-dependent attenuation. The goal of this module is to learn discrete latent representations that preserve essential physical propagation patterns while reducing redundancy.
The input pathloss map is a single-channel image $\mathit{P} \in \mathbf{R}^{H \times W \times 1}$, where each pixel denotes the signal attenuation value at a specific Tx-Rx location. To efficiently encode the pathloss map in electromagnetic space, a ResNet-based VQGAN is employed to extract and quantize electromagnetic propagation features while maintaining spatial continuity. The input pathloss map first passes through an encoder $E_{\mathrm{C}}$ composed of convolutional and residual blocks, generating a continuous latent representation, which is written as
\begin{equation}
\hat{z}^{C} = E_{\mathrm{C}}(P) \in \mathbf{R}^{h' \times w' \times n_z'}
\end{equation}
where $h' \times w' \times n_z'$ denotes the dimensionality of the continuous latent representation vector.
A discrete codebook $\mathcal{Z'} = \{{ z_k' }\}_{k'=1}^{K'}, z_k' \in \mathbf{R}^{n_z'}$ is pre-defined, and each feature vector in $\hat{z}^{C}$ is replaced by its nearest codeword $z_q^{C}$ through a quantization operation $\mathrm{Q}$, which is presented as
\begin{equation}
z_q^{C} = \mathrm{Q}(\hat{z}^{C}, \mathcal{Z'}).
\end{equation}
The resulting $z_q^{C}$ serves as the encoded representation of the pathloss map in the channel synesthetic space.
During training, the VQGAN is optimized using a hybrid loss $\mathcal{L}$ combining vector quantization and adversarial terms, which is presented as
\begin{equation}
\mathcal{L'} = \mathcal{L'}_{\mathrm{VQ}} + \lambda' \mathcal{L'}_{\mathrm{GAN}}
\end{equation}
where
\begin{equation}
\begin{aligned}
\mathcal{L'}_{\mathrm{VQ}} = || P - \hat{P} ||^2 + 
|| \mathrm{sg}[E_{\mathrm{C}}(P)] - z_q^{C} ||^2 \\ + || \mathrm{sg}(z_q^{C}) - E_{\mathrm{C}}(P) ||^2
\end{aligned}
\end{equation}
and
\begin{equation}
\begin{aligned}
\mathcal{L'}_{\mathrm{GAN}}
= &  \log D'(P) + \log \big( 1 - D'(\hat{P}) \big).
\end{aligned}
\end{equation}
In (10) and (11), $\hat{P}$ denotes the reconstructed pathloss map, $\mathrm{sg}[\cdot]$ represents the stop-gradient operation, $\lambda'$ is a weighting coefficient for the adversarial term, and $D'$ denotes the discriminator network, which is trained to distinguish between the real pathloss map $P$ and the reconstructed one $\hat{P}$. The discriminator adopts a hinge loss formulation to improve the perceptual realism of reconstructed maps, particularly in preserving spatial gradients and frequency-dependent variations.
Through this discrete encoding process, pathloss maps are transformed into compact and interpretable latent codes in the channel synesthetic space, forming the basis for cross-modal alignment and generation with the sensory synesthetic space.

\subsection{Stage 2: Mapping Mechanism Exploration}
\begin{figure}
\centerline{\includegraphics[width=0.45\textwidth]{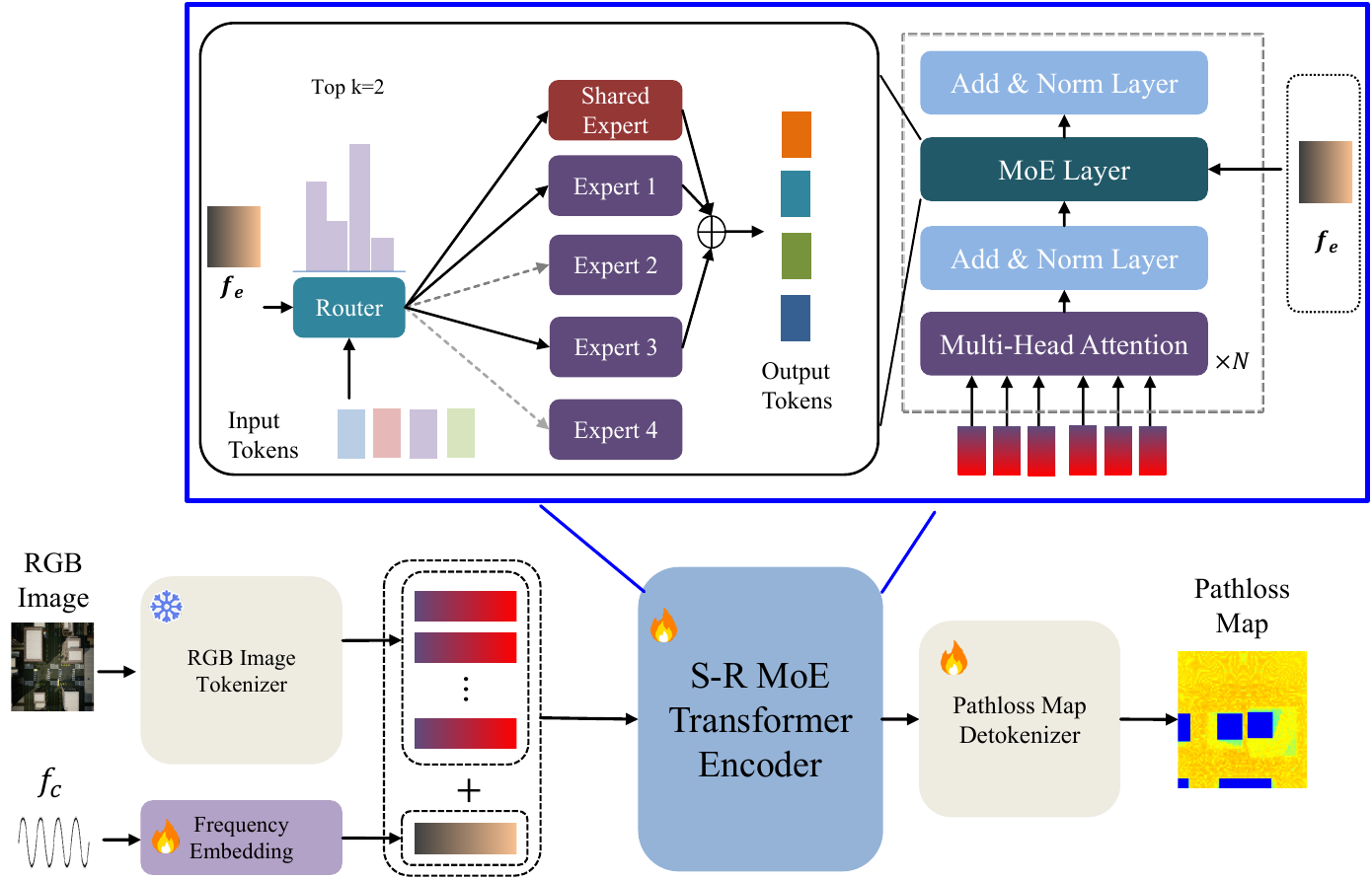}}
\caption{An illustration of the mapping mechanism exploration module.\label{moe}}
\end{figure}

In the stage 2, the mapping mechanism exploration module is developed, where a frequency-guided S-R MoE Transformer module is constructed to achieve cross-modal mapping from the sensory synesthetic space to the channel synesthetic space. Based on the designed frequency-guided S-R MoE Transformer, the model enables cross-modal generation and generalization from RGB images to pathloss maps at various conditions, as illustrated in Fig.~\ref{moe}. 

For different communication frequency bands, a joint frequency embedding method is proposed, consisting of two parts, i.e., frequency ID embedding and frequency value embedding. The frequency ID embedding module assigns a unique identifier to each frequency ID $f_{\mathrm{id}}$ and maps it to a fixed-dimensional vector $e_{\mathrm{id}}$ via a learnable embedding network ${E}$, which is presented as 
\begin{equation}
    e_{\mathrm{id}} = {E}(f_{\mathrm{id}}) \in \mathbb{R}^{d}
\end{equation}
where $d$ denotes the embedding dimension.  
To incorporate the frequency information of each frequency band into the model, a frequency band sinusoidal–cosine encoding mechanism is employed. For a given frequency value $f \in \mathbf{R}$, the encoding module projects it into a $d$-dimensional embedding vector by generating sinusoidal and cosine components with exponentially increasing scales.
Formally, the encoding function $\mathrm{E_f}(\cdot)$ is defined as
\begin{equation}
\begin{aligned}
\mathrm{E_f}(f) = 
\big[ & \sin(2\pi f), \, \sin(2\pi f \cdot 2^{1}), \, \dots, \,
\sin(2\pi f \cdot 2^{D-1}), \\
& \cos(2\pi f), \, \cos(2\pi f \cdot 2^{1}), \, \dots, \,
\cos(2\pi f \cdot 2^{D-1}) \big] \\
\end{aligned}
\end{equation}
where $D = d/2$ is the number of sinusoidal bases.
Each dimension of the embedding corresponds to a sinusoidal function with a frequency base that grows exponentially as $2^{i}$, which allows the model to capture variations in frequency bands. Given a batch of frequency values $\{{ f_i\} }_{i=1}^{B}$, the resulting encoding matrix ${e_f} \in \mathbf{R}^{B \times d}$ is concatenated as
\begin{equation}
{e_f} = [\sin(2\pi f \cdot 2^{i}) , ; , \cos(2\pi f \cdot 2^{i})], \quad i = 0, 1, \dots, D-1.
\end{equation}
The aforementioned encoding provides a smooth and continuous representation of frequency variations, enabling the model to learn frequency-dependent propagation characteristics in the cross-modal pathloss generation.
Finally, the two embeddings are concatenated to form the frequency-guided vector, which is written as
\begin{equation}
    e_f' = \mathrm{concat}[e_{\mathrm{id}},\ e_{{f}}].
\end{equation}
The aforementioned vector $e_f'$ preserves both the physical meaning of the frequency and the learnable embedding capability, serving as the conditional input for the subsequent S-R MoE gating mechanism.

Then, a frequency-guided S-R MoE Transformer is designed and implemented to perform cross-modal mapping from the perceptual synesthetic space to the channel synesthetic space.  
Unlike conventional single-path Transformers, the proposed architecture introduces a S-R MoE architecture, where a frequency-guided dynamic gating network adaptively activates specialized experts for different frequency inputs. This mechanism enhances both the accuracy and robustness of pathloss generation across diverse frequency conditions. The discrete latent token sequence in the sensory synesthetic space is denoted as $z_q^{S}$, obtained from the RGB image encoder. Meanwhile, the communication frequency $f$ is input into the frequency embedding module to produce the fused frequency vector $e_f'$.

The S-R MoE Transformer module consists of multiple stacked Transformer blocks, each containing one shared–routing substructure composed of one shared expert $E_s(\cdot)$ and multiple routing experts $\{E_r^1, E_r^2, \ldots, E_r^{N_r}\}$.  
The shared expert $E_s(\cdot)$ learns frequency-independent cross-modal common features, while the routing experts capture frequency-specific pathloss characteristics. Each expert is implemented as an independent feed-forward subnetwork.  
A Top-2 routing strategy is adopted, where only the two routing experts with the highest weights determined by $e_f'$ are activated, thus balancing model expressiveness and computational efficiency.  
The layer output is obtained by weighted fusion of the shared expert and the activated routing experts, which is calculated as
\begin{equation}
    Y_l = \alpha_s E_s(z_q^{s}) + \sum_{j \in \mathrm{Top\text{-}2}} {\alpha_j^{e_f'} E_r^j(z_q^{s})}
\end{equation}
where $z_q^{s}$ represents the input of the feed forward layer, $\alpha_s$ and $\alpha_j^{e_f'}$ represent the weighting coefficients corresponding to the outputs of different experts, and $Y_l$ represent the output of the current Transformer block, which is mapped to the channel synesthetic space in the final block. This mechanism enables the model to dynamically select appropriate expert paths according to the input frequency, achieving frequency-adaptive structural modulation in cross-modal pathloss generation.

\section{Simulation Results and Analysis}

In this section, the simulation results and analysis of the proposed WiCo-PG are shown. First, the setup of experiments for model training and testing is introduced. Furthermore, the full-sample pathloss map generation performances under single condition and multi-condition dataset unified learning are illustrated, respectively. Moreover, the generalization performances across different scenarios, flight altitudes, and frequency bands of the proposed WiCo-PG are shown. Finally, the results of the scaling analysis and the ablation experiments are presented.

\subsection{Setup}
The detailed hyper-parameters of the WiCo-PG network design and pre-training are listed in Table \ref{setup}. The Transformer module is configured with 16 attention heads and an embedding dimension of 256 for each input token. A total of 8 Transformer blocks are stacked to perform frequency-guided cross-modal mapping. For the VQGAN, the size of the discrete codebook is set to 2048 vectors to balance representation richness and computational efficiency. During pre-training, both the VQGAN and Transformer modules are optimized using the ADAM optimizer \cite{adam} with learning rates of $2.25 \times 10^{-5}$ and $4.45 \times 10^{-5}$, respectively. The model is trained for 500 epochs with a batch size of 16 to ensure stable convergence and robust generalization performance.

\begin{table}[t!]
\renewcommand\arraystretch{1.5}
\centering
\caption{Hyper-Parameter for WiCo-PG Design and Pre-Training.}
\begin{tabular}{|c|c|}
\hline
\textbf{Parameter} & \textbf{Value} \\ \hline
Number of attention heads & 16 \\ \hline
Embedding dimension of input tokens & 256 \\ \hline
Number of codebook vectors & 2048 \\ \hline
Number of Transformer blocks & 8 \\ \hline
Batch size & 16 \\ \hline
Learning rate of the VQGAN & $2.25 \times 10^{-5}$ \\ \hline
Learning rate of the Transformer & $4.45 \times 10^{-5}$ \\ \hline
Epochs & 500 \\ \hline
Optimizer & ADAM \\ \hline

\end{tabular}
\label{setup}
\end{table}

For comparison, two baseline models are employed to evaluate the pathloss map generation performance. The first baseline is a conventional deep learning artificial intelligence generated content (AIGC) model based on a GAN. The generator in this model adopts a ResNet architecture to capture hierarchical visual features, while the discriminator is implemented as a CNN network to distinguish between real and generated pathloss maps. The second baseline is the LLM-based pathloss map generation model, denoted as LLM4PG \cite{LLM4PG}. This model adapts a pre-trained GPT-2 model through fine-tuning to perform cross-modal pathloss map generation. The aforementioned two models are selected as baselines because they represent two typical approaches for pathloss map generation. The GAN-based AIGC model serves as a representative of conventional deep learning approaches that rely on visual feature learning, while LLM4PG exemplifies the approaches of leveraging LLMs for cross-modal pathloss map generation. Comparing with both models allows a comprehensive evaluation of the proposed WiCo-PG in terms of generation accuracy, modality understanding, and generalization capability. The performance comparison with LLM4PG and the GAN-based model demonstrates the advantage of the proposed WiCo-PG in achieving improved generation accuracy and enhanced generalization across various conditions.

\subsection{Full-Sample Pathloss Map Generation Performance of WiCo-PG}

\begin{figure}
\centerline{\includegraphics[width=0.45\textwidth]{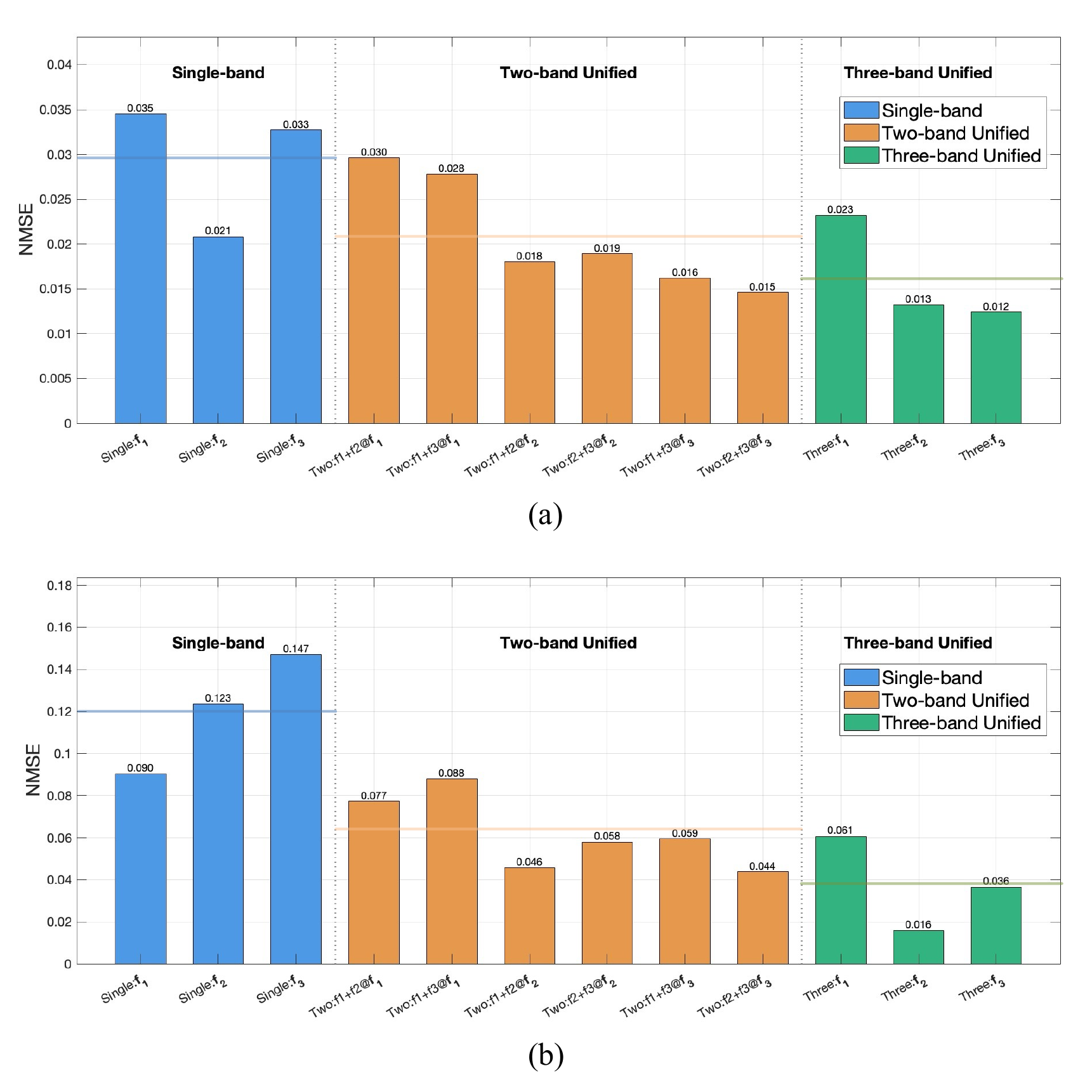}}
\caption{The NMSE performance of the proposed WiCo-PG under multi-band dataset unified learning. (a) Urban crossroad scenario. (b) Urban wide lane scenario.\label{multi-freq}}
\end{figure}

\begin{figure}
\centerline{\includegraphics[width=0.45\textwidth]{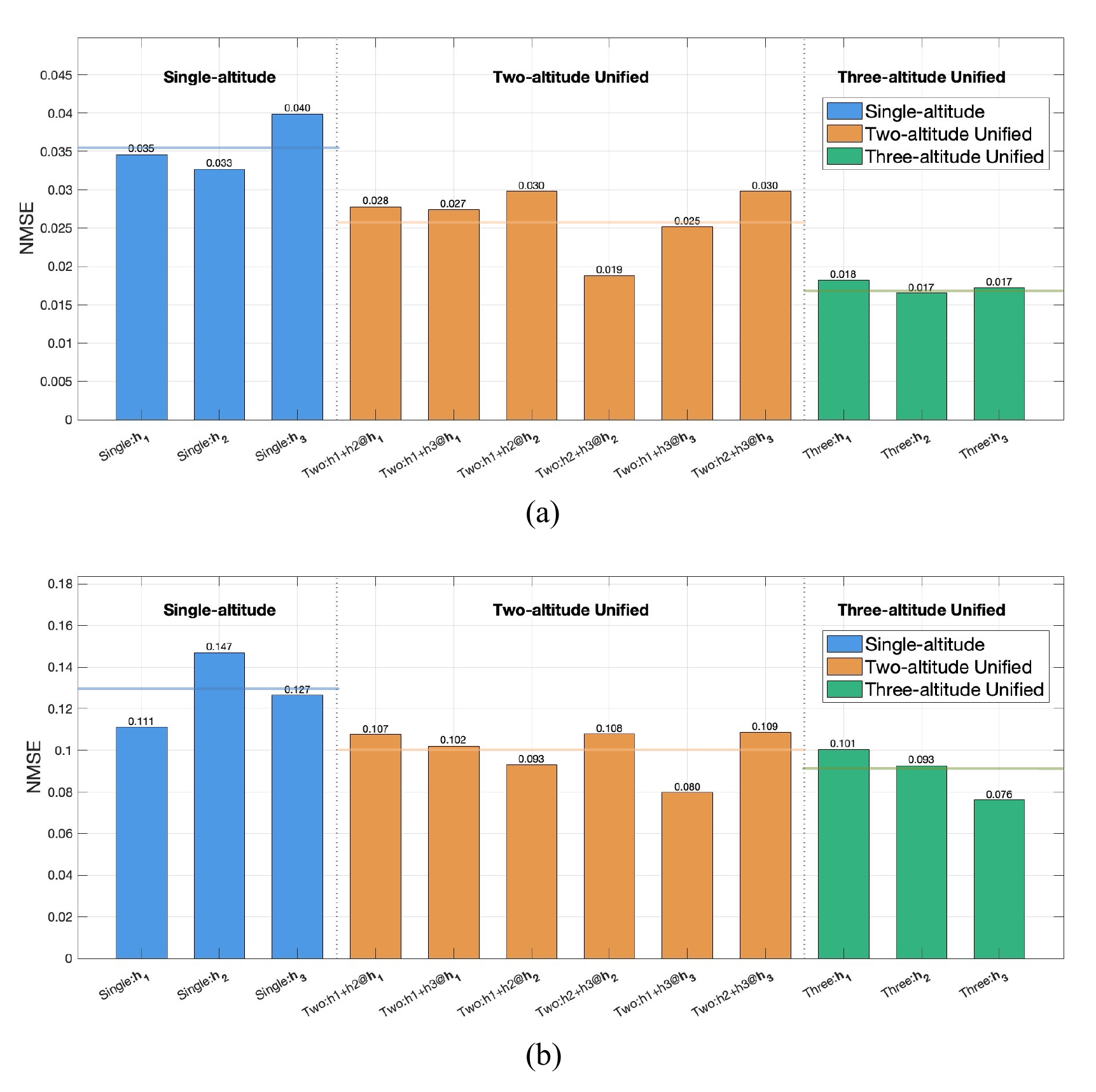}}
\caption{The NMSE performance of the proposed WiCo-PG under multi-altitude dataset unified learning. (a) Urban crossroad scenario. (b) Urban wide lane scenario.\label{multi-alti}}
\end{figure}

The full-sample pathloss map generation performances of the WiCo-PG under a single condition and multi-condition dataset unified learning are illustrated. For the full-sample pathloss map generation performance, the NMSE of pathloss map generation is evaluated across diverse frequency bands and diverse altitudes within different scenarios. The pathloss map generation performance on a specified dataset is evaluated using the average NMSE, which can be calculated as

\begin{equation}
\label{deqn_ex1}
\text{NMSE} = \mathbb{E} \left\{ 
\frac{
\sum_{i=1}^{N_t} \left\| \hat{\mathbf{P}}_\mathbf{M} - \mathbf{P_M} \right\|^2
}{
\sum_{i=1}^{N_t} \left\| \mathbf{P_M} \right\|^2
}
\right\}
\end{equation}
where $\mathbf{P_M}$ denotes the ground truth obtained from RT simulation and ${\mathbf{\hat{P}_M}}$ denotes the pathloss map generated by the proposed WiCo-PG. $\mathbb{E}[\cdot]$ indicates the statistical expectation taken over all samples in the specified test set. The subscript $i$ refers to the index of the $i$-th sample among the total of $N_t$ test samples. $\left| \cdot \right|^2$ represents the squared Frobenius norm, which quantifies the element-wise squared error between two matrices.

To demonstrate the capability of the proposed WiCo-PG in multi-condition dataset unified learning, we conduct unified training across datasets with diverse frequency bands and diverse flight altitudes in different scenarios, and compare the results with the full-sample performance under a single condition. Specifically, as shown in Fig.~\ref{multi-freq}, the pathloss map generation performance of multi-band dataset unified training and full-sample pathloss map generation performance under single condition in different scenarios are presented. As shown in Fig.~\ref{multi-alti}, the pathloss map generation performance of multi-altitude dataset unified training and full-sample pathloss map generation performance under single condition in different scenarios are presented. The blue bars in Figs.~\ref{multi-freq} and~\ref{multi-alti} illustrate the full-sample training performance under single condition at different frequency bands, i.e., \( f_1 = 1.6\,\mathrm{GHz} \), \( f_2 = 15\,\mathrm{GHz} \), and \( f_3 = 28\,\mathrm{GHz} \), or at different flight altitudes, i.e., \( h_1 = 50\,\mathrm{m} \), \( h_2 = 70\,\mathrm{m} \), and \( h_3 = 80\,\mathrm{m} \) in the urban crossroad scenario and \( h_1 = 200\,\mathrm{m} \), \( h_2 = 250\,\mathrm{m} \), and \( h_3 = 300\,\mathrm{m} \) in the urban wide lane scenario. The orange bars in Figs.~\ref{multi-freq} and~\ref{multi-alti} illustrate the results of multi-condition dataset unified training on two frequency bands or two flight altitudes. \( f_i + f_j @ f_i \) denotes that the proposed WiCo-PG is trained on datasets of frequency bands \( f_i \) and \( f_j \), and then tested on the frequency band \( f_i \). \( h_i + h_j @ h_i \) denotes that the proposed WiCo-PG is trained on datasets of flight altitudes \( h_i \) and \( h_j \), and then tested on the altitude \( h_i \). The green bars in Figs.~\ref{multi-freq} and~\ref{multi-alti} illustrate the test results on each frequency band or flight altitude after multi-condition dataset unified training across all three frequency bands or flight altitudes.

For the multi-band dataset unified learning, in the urban crossroad scenario, datasets with diverse frequency bands are collected at the same flight altitude of 70\,m. In the urban wide lane scenario, datasets with diverse frequency bands are collected at the same flight altitude of 200\,m. As shown in Fig.~\ref{multi-freq}(a), in urban crossroad scenario, the average NMSE of the full-sample training on a single frequency band is 0.030, while the average NMSE of testing on each frequency band after multi-condition dataset unified learning with two frequency bands is 0.021. Furthermore, when unified learning is performed across three frequency bands, the average NMSE of testing on each frequency band decreases to 0.016. As shown in Fig.~\ref{multi-freq}(b), in urban wide lane scenario, the average NMSE of the full-sample training on a single frequency band is 0.120, while the average NMSE of testing on each frequency band after multi-condition dataset unified learning with two frequency bands is 0.062. Furthermore, when unified learning is performed across three frequency bands, the average NMSE of testing on each frequency band decreases to 0.038. In both the urban crossroad and urban wide lane scenarios, the average NMSE of pathloss map generation decreases as the number of frequency-band datasets increases, where the average NMSE is indicated by the horizontal lines between the bars in the figures. Moreover, for the test results on the same frequency band, the pathloss map generation performance gradually improves with the increasing number of frequency-band datasets. This phenomenon can be attributed to the designed frequency-guided S-R MoE Transformer, which adaptively leverages shared information across bands while preserving frequency-specific distinctions, allowing the model to achieve more effective feature fusion and thus improving the overall performance through multi-band dataset unified learning.

For the multi-altitude dataset unified learning, in the urban crossroad scenario, datasets of different flight altitudes are collected at the same frequency band of \( f = 28\,\mathrm{GHz} \), while in the urban wide lane scenario, datasets of different flight altitudes are collected at the same frequency band of \( f = 28\,\mathrm{GHz} \). As shown in Fig.~\ref{multi-alti}(a), in the urban crossroad scenario, the average NMSE of the full-sample training at a single altitude is 0.036, while the average NMSE of testing on each altitude after unified learning with two altitudes is 0.027. Furthermore, when unified learning is performed across three altitudes, the average NMSE of testing on each altitude decreases to 0.017. As shown in Fig.~\ref{multi-alti}(b), in the urban wide lane scenario, the average NMSE of the full-sample training at a single altitude is 0.128, while the average NMSE of testing on each altitude after unified learning with two altitudes is 0.100. Moreover, when unified learning is performed across three altitudes, the average NMSE of testing on each flight altitude further decreases to 0.090. In both scenarios, the average NMSE of pathloss map generation gradually decreases as the number of flight-altitude datasets increases, as indicated by the horizontal lines between the bars in the figures. Meanwhile, for tests conducted at the same altitude, the pathloss map generation performance consistently improves with the inclusion of more flight altitude-diverse datasets. This phenomenon arises from the fact that the multi-altitude unified learning enables the WiCo-PG to capture more comprehensive spatial–semantic representations of the propagation environment, allowing it to better understand altitude-dependent variations in spatial distribution of pathloss and thus achieve improved overall performance.

In summary, the performance patterns exhibited by the proposed WiCo-PG in multi-condition dataset unified learning demonstrate that the proposed WiCo-PG is capable of unified learning across different flight altitudes and frequency bands during the pre-training stage. Meanwhile, the results also validate the effectiveness of the designed frequency-guided S-R MoE architecture. As a result, the proposed WiCo-PG can accurately generate pathloss maps under various input conditions by adapting to RGB image inputs from different altitudes and carrier frequency inputs from different frequency bands.

\subsection{Pathloss Map Generation Performance of WiCo-PG Across Diverse Flight Altitudes}

This subsection presents the cross-condition generalization results of the proposed WiCo-PG across diverse flight altitudes.  To verify the generalization capability of the proposed WiCo-PG, we evaluate the pathloss map generation performance under zero-shot generalization and few-shot fine-tuning generalization. Specifically, zero-shot generalization refers to the case where the WiCo-PG is trained under specific conditions and directly tested under another condition without any fine-tuning. Few-shot fine-tuning generalization refers to the case where the WiCo-PG is trained under a specific condition, fine-tuned with a few samples from another condition, and then tested under that new condition. For the pathloss map generation performance of WiCo-PG across diverse flight altitudes, generalization experiments across different flight altitudes are conducted in the urban crossroad scenario, where the WiCo-PG is trained with unified datasets from two flight altitudes and then evaluated on its zero-shot and few-shot fine-tuning generalization capabilities at another unseen flight altitude.

\begin{figure}
\centerline{\includegraphics[width=0.48\textwidth]{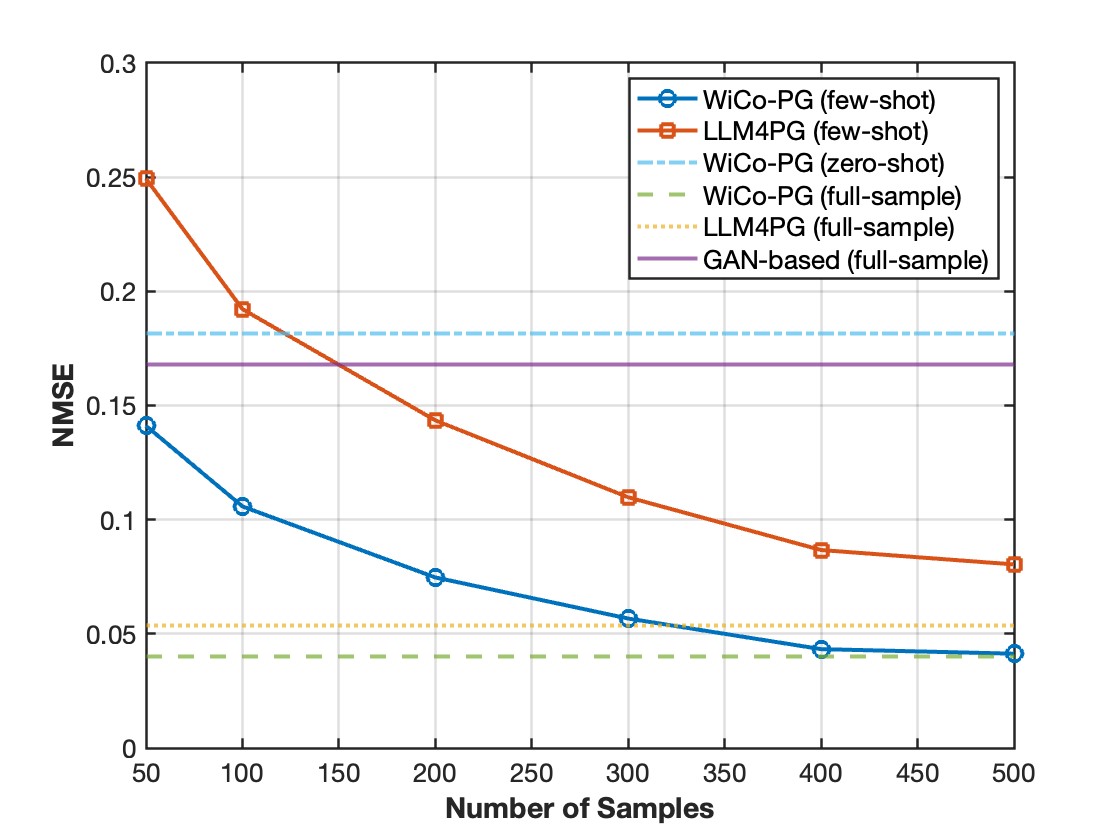}}
\caption{The cross-flight altitude generalization performance from 50\,m and 70\,m to unseen 80\,m.\label{50+70-80}}
\end{figure}

\begin{figure}
\centerline{\includegraphics[width=0.48\textwidth]{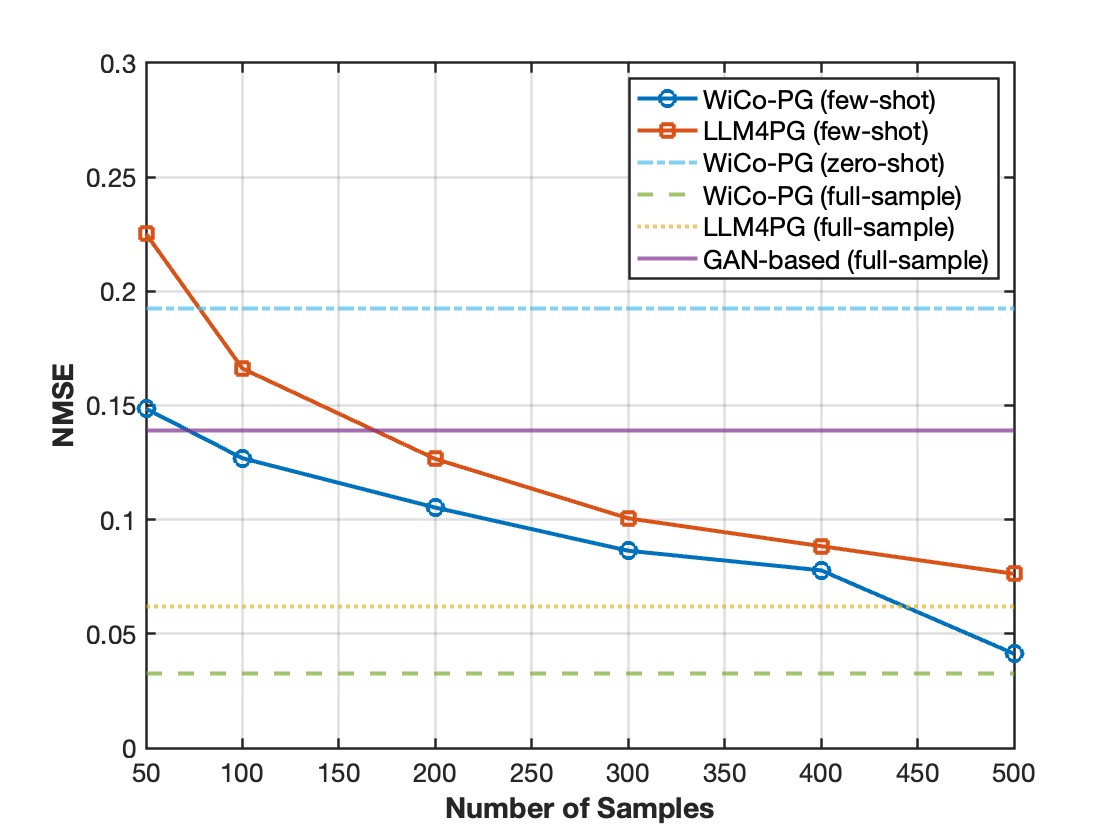}}
\caption{The cross-flight altitude generalization performance from 50\,m and 80\,m to unseen 70\,m.\label{50+80-70}}
\end{figure}

\begin{figure}
\centerline{\includegraphics[width=0.48\textwidth]{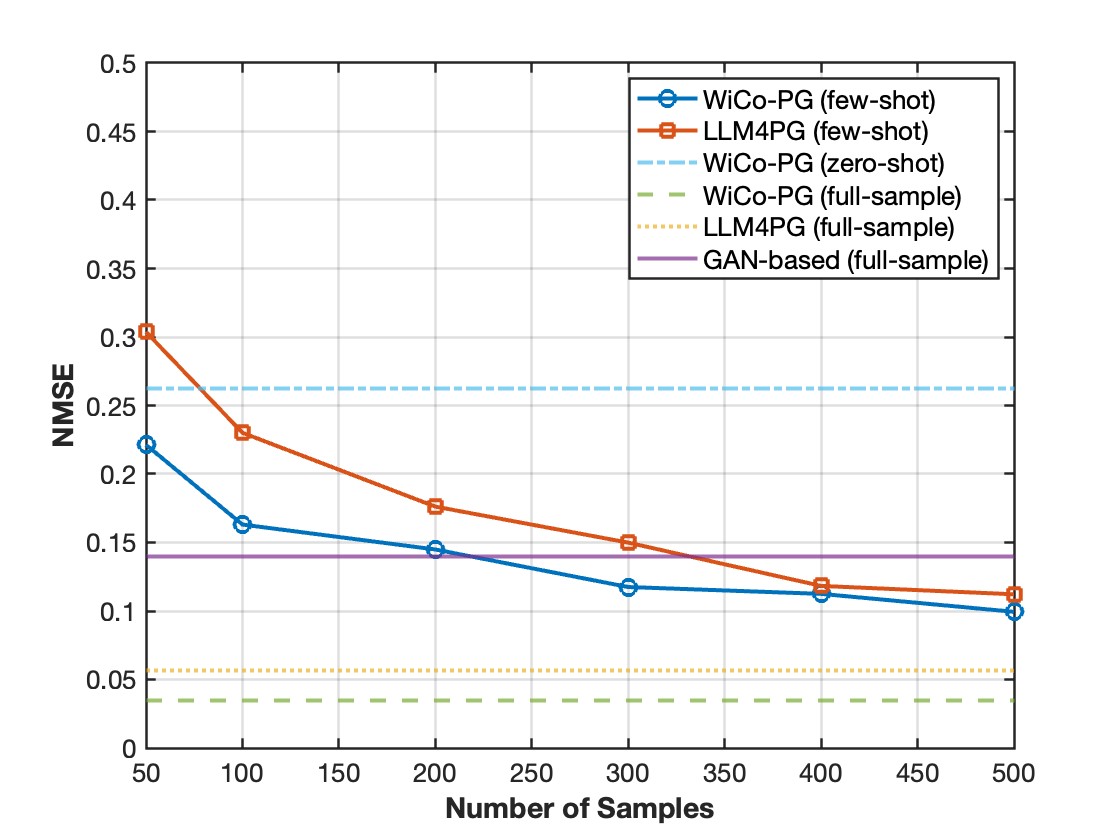}}
\caption{The cross-flight altitude generalization performance from 70\,m and 80\,m to unseen 50\,m.\label{70+80-50}}
\end{figure}

Figs.~\ref{50+70-80}--\ref{70+80-50} present the cross-flight altitude generalization results of the proposed WiCo-PG, along with comparisons to the LLM4PG scheme and the GAN-based model. Specifically, Fig.~\ref{50+70-80} illustrates the generalization results when the model is trained at multi-altitude datasets of 50\,m and 70\,m, and evaluated at the unseen altitude of 80\,m. Fig.~\ref{50+80-70} illustrates the generalization results when the model is trained at multi-altitude datasets of 50\,m and 80\,m, and evaluated at the unseen altitude of 70\,m. Fig.~\ref{70+80-50} illustrates the generalization results when the model is trained at multi-altitude datasets of 70\,m and 80\,m, and evaluated at the unseen altitude of 50\,m. The results indicate that the proposed WiCo-PG possesses zero-shot generalization capability across different flight altitudes, achieving an NMSE of 0.182. Moreover, in the few-shot generalization, the proposed WiCo-PG achieves an improvement of over 1.37\,dB compared with the LLM4PG using no more than 2.7\% of samples, and an improvement of over 0.53\,dB using no more than 27.3\% of samples. The superior cross-altitude generalization of WiCo-PG is attributed to its ability to capture altitude-dependent spatial–semantic relationships between the physical environment and signal propagation, allowing the WiCo-PG to effectively adapt to changes in observation perspective and maintain more stable pathloss map generation across different flight altitudes than other models. These results demonstrate that the proposed WiCo-PG has the potential to enable massive and high-quality pathloss data generation across diverse flight altitudes. In addition, generalizing from lower to higher flight altitudes is relatively easier than the reverse, since lower-altitude conditions are often affected by severe signal blockage and complex multipath effects caused by surrounding buildings and obstacles \cite{blockage-twc}, resulting in more intricate pathloss patterns. Consequently, models trained under these more challenging low-altitude conditions can better generalize to higher-altitude scenarios, where the propagation environment becomes simpler with fewer blockages and more stable pathloss distributions.

\subsection{Pathloss Map Generation Performance of WiCo-PG Across Different Scenarios}

\begin{figure}
\centerline{\includegraphics[width=0.48\textwidth]{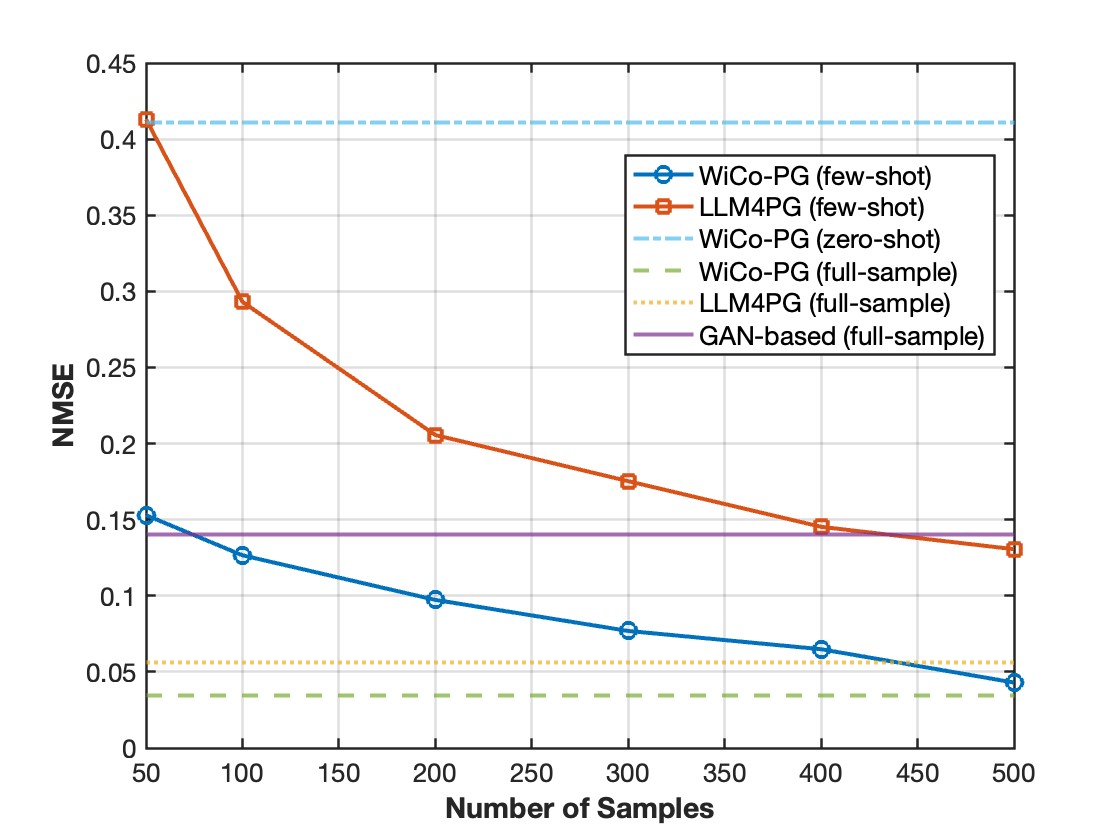}}
\caption{The cross-scenario generalization performance from  wide lane scenario to unseen crossroad scenario.\label{ck-sz}}
\end{figure}

This subsection presents the cross-scenario generalization results of the proposed WiCo-PG. Similar to the cross-flight altitude generalization experiments, the cross-scenario generalization experiments evaluate the zero-shot generalization and few-shot fine-tuning performance of the proposed WiCo-PG. Specifically, as shown in Fig.~\ref{ck-sz}, the model is first trained in the urban crossroad scenario and then tested in the urban wide lane scenario. The results indicate that the proposed WiCo-PG possesses zero-shot generalization capability across different scenarios, achieving an NMSE of 0.41. Moreover, in the few-shot generalization, the proposed WiCo-PG achieves an improvement of over 4.82\,dB compared with the LLM4PG using no more than 2.7\% of samples, and an improvement of over 4.31\,dB using no more than 27.3\% of samples. The enhanced cross-scenario generalization of WiCo-PG is attributed to its ability to capture shared spatial–semantic patterns across diverse physical environments while adapting to scenario-specific differences, enabling more accurate and stable pathloss map generation under varying scenarios than other models. This demonstrates the excellent cross-scenario generalization ability of WiCo-PG and its potential for massive and high-quality pathloss map generation and generalization across diverse scenarios.

\subsection{Pathloss Map Generation Performance of WiCo-PG Across Diverse Frequency Bands}

\begin{figure}
\centerline{\includegraphics[width=0.48\textwidth]{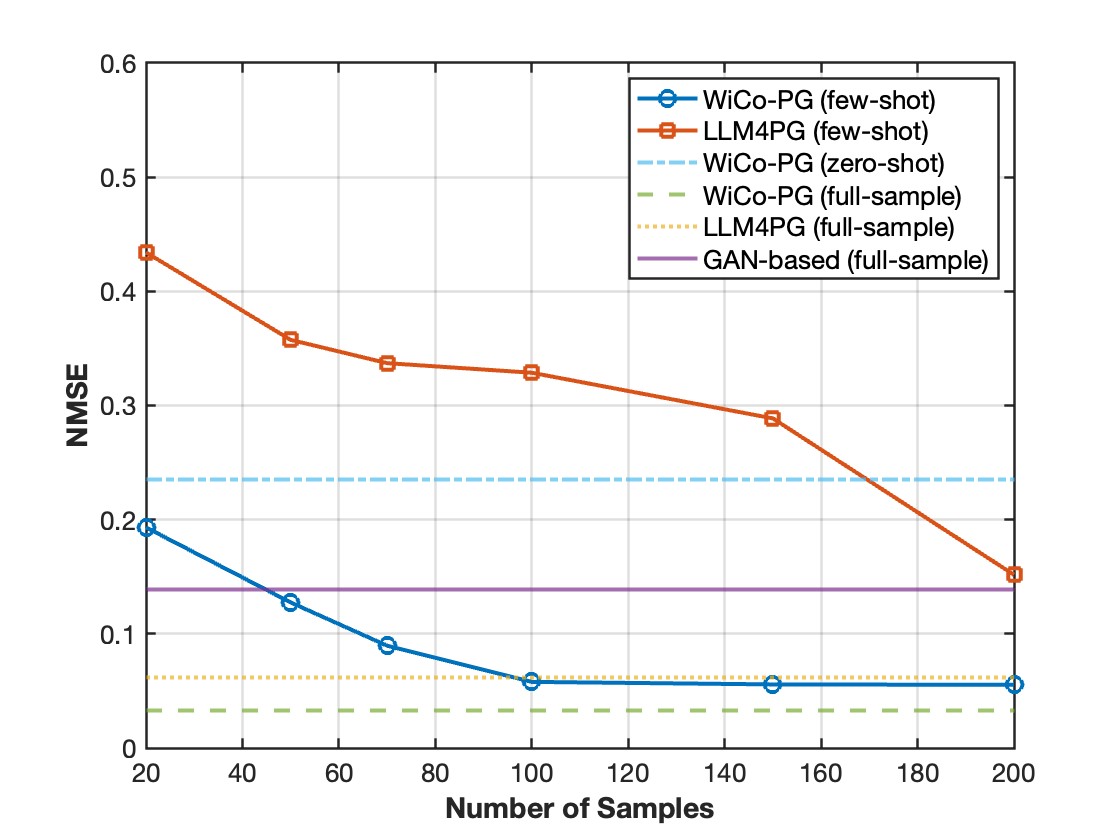}}
\caption{The cross-frequency band generalization performance from 1.6\,GHz and 15\,GHz to unseen 28\,GHz.\label{sub6+15-28}}
\end{figure}

\begin{figure}
\centerline{\includegraphics[width=0.48\textwidth]{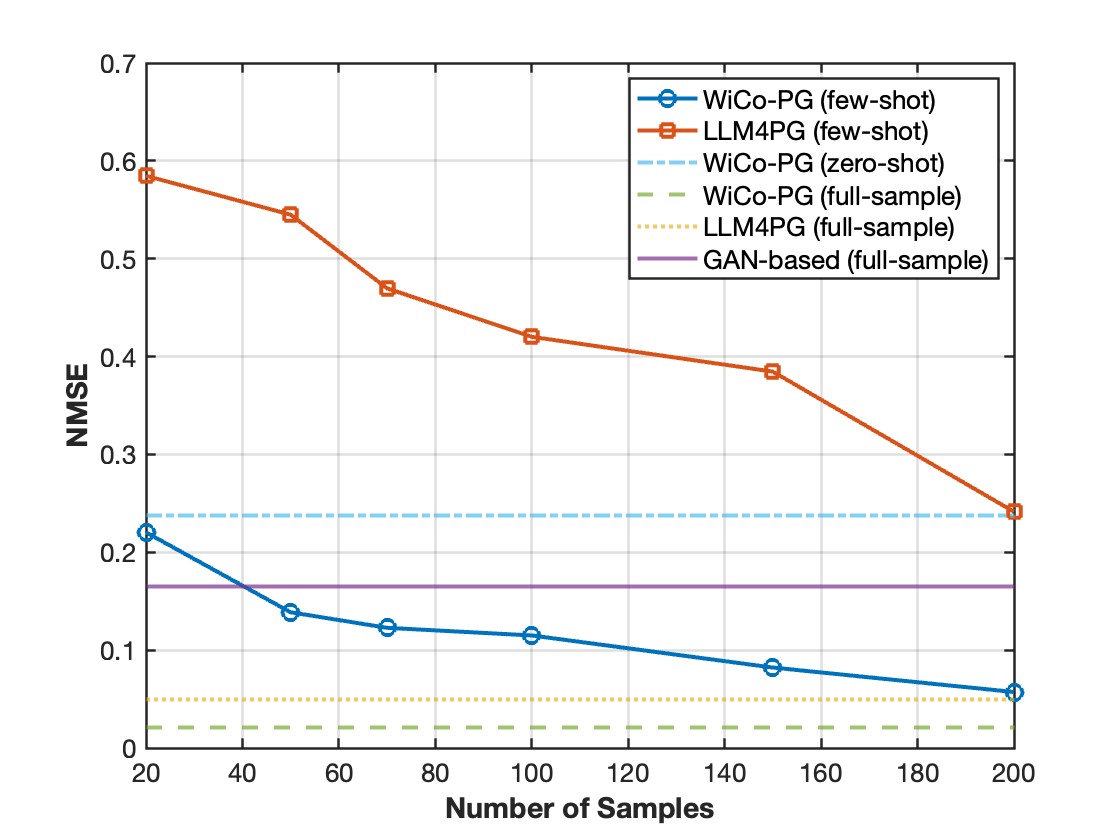}}
\caption{The cross-frequency band generalization performance from 1.6\,GHz and 28\,GHz to unseen 15\,GHz.\label{sub6+28-15}}
\end{figure}

\begin{figure}
\centerline{\includegraphics[width=0.48\textwidth]{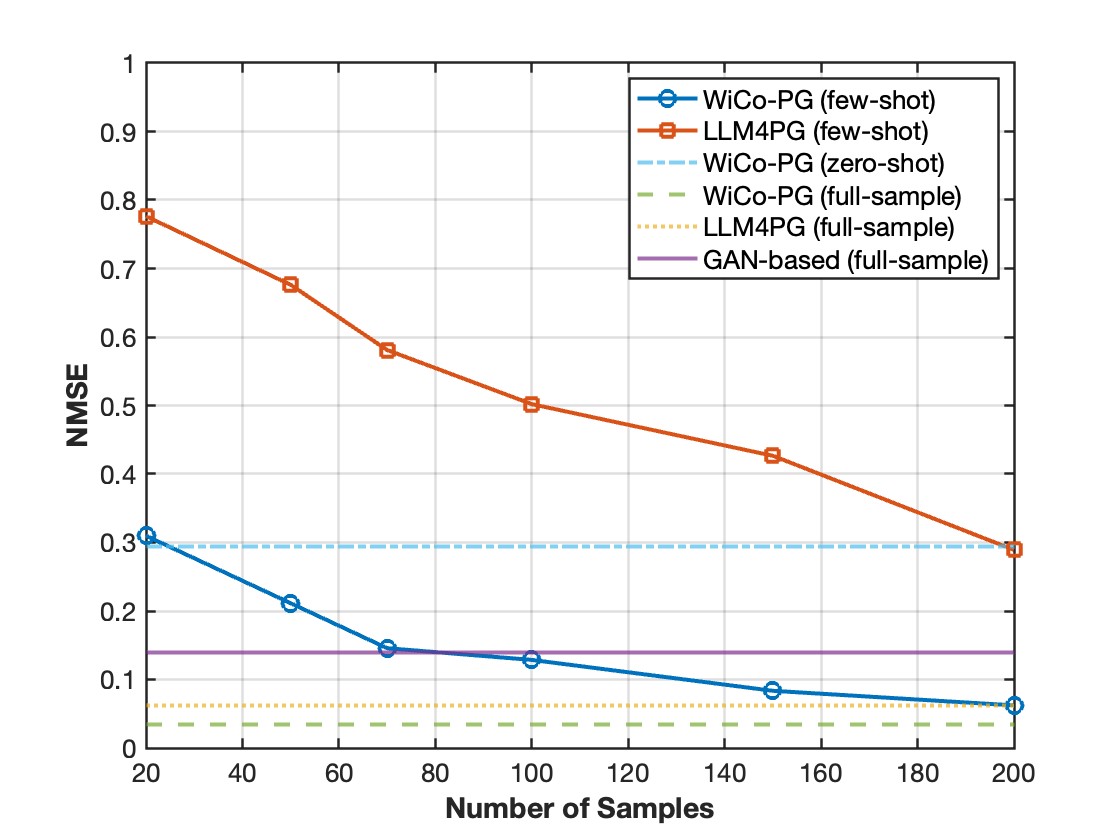}}
\caption{The cross-frequency band generalization performance from 15\,GHz and 28\,GHz to unseen 1.6\,GHz.\label{15+28-sub6}}
\end{figure}

This subsection presents the generalization capability of the proposed WiCo-PG across different frequency bands. Similar to the cross-flight altitude generalization experiments, the model trained with multi-band datasets from two frequency bands is evaluated on the unseen frequency band for zero-shot generalization and few-shot fine-tuning performance.
Figs.~\ref{sub6+15-28}--\ref{15+28-sub6} present the cross-frequency band generalization results of the proposed WiCo-PG, along with comparisons to the LLM4PG and the GAN-based model. Specifically, Fig.~\ref{sub6+15-28} illustrates the generalization results when the model is trained at multi-band datasets of 1.6\,GHz and 15\,GHz, and evaluated at the unseen frequency band of 28\,GHz. Fig.~\ref{sub6+28-15} illustrates the generalization results when the model is trained at multi-band datasets of 1.6\,GHz and 28\,GHz, and evaluated at the unseen frequency band of 15\,GHz. Fig.~\ref{15+28-sub6} illustrates the generalization results when the model is trained at multi-band datasets of 15\,GHz and 28\,GHz, and evaluated at the unseen frequency band of 1.6\,GHz.
The results indicate that the proposed WiCo-PG possesses zero-shot generalization capability across different frequency bands, achieving an NMSE of 0.24. Moreover, in the few-shot generalization, The proposed WiCo-PG achieves an improvement of over 1.37\,dB compared with LLM4PG when using no more than 1.1\% of the samples, and an improvement of over 3.52\,dB when using no more than 10.9\% of the samples. The superior cross-band generalization of WiCo-PG benefits from its frequency-guided S-R MoE Transformer and pre-training strategy, which enable adaptive modeling of frequency-dependent propagation characteristics while leveraging shared spatial–semantic representations across bands, leading to more robust pathloss map generalization capability under varying bands than other models. These results demonstrate that the proposed WiCo-PG has the potential to enable massive and high-quality pathloss data generation across diverse frequency bands. In addition, generalizing from lower to higher frequency bands is relatively easier than the reverse, as lower-band conditions exhibit richer scattering and more complex propagation characteristics, making generalization toward higher-band conditions less challenging.

\subsection{Scaling Analysis and Ablation Study}

This subsection presents the scaling analysis and ablation study, which examine the effects of different dataset scales and model sizes on the performance of the proposed foundation model WiCo-PG, as well as the impact of the designed modules on the overall experimental results. Specifically, for the scaling analysis, the effects of different dataset scales on model performance under the same model parameters are evaluated, as well as the effects of varying model parameters on performance under the same dataset scale. As shown in Table~\ref{scale}, the results show that, consistent with the conclusions in Section IV-B, the model achieves better performance with larger dataset scales under the same number of model parameters. D1 represents the dataset collected in the crossroad scenario at a flight altitude of 70 m and a carrier frequency of 1.6\,GHz, D2 corresponds to the dataset in the same scenario and altitude at 15\,GHz, and D3 denotes the data at 28\,GHz under identical conditions. In addition, for the same dataset scale, a larger number of model parameters leads to stronger learning and representation capabilities, resulting in improved performance. The inference time is tested on the AMD Ryzen Threadripper PRO 7955WX 16-Core CPU, the NVIDIA GeForce RTX4090 GPU, and 128\,GB of RAM.  For the ablation study, as shown in Table~\ref{ablation}, the performance variations are evaluated on the same dataset with and without the proposed architectural design. The results demonstrate that the frequency-guided S-R MoE architecture plays a crucial role in enabling WiCo-PG to achieve high-accuracy and zero-shot generalization capabilities, facilitating massive and high-quality data generation under diverse conditions.

\begin{table}[h!]
\renewcommand\arraystretch{1.5}
\setlength{\tabcolsep}{4pt} 
\small 
\centering
\caption{NMSE Performance of WiCo-PG under Different Dataset Combinations and Model Parameters.}
\begin{tabular}{|c|c|c|c|c|}
\hline
\textbf{Dataset} & \textbf{Model} & \textbf{NMSE} & \textbf{Params} & \textbf{Infer. time} \\ \hline

\multirow{3}{*}{D1, D2, D3}
& WiCo-PG-small & 0.0596 & 145.6\,M & 14.3\,ms \\ \cline{2-5}
& WiCo-PG-large & \textbf{0.0127} & 519.6\,M & 17.1\,ms \\ \cline{2-5}
& WiCo-PG-base  & \underline{0.0163} & \multirow{4}{*}{169.7\,M} & \multirow{4}{*}{16.3\,ms} \\ \cline{1-3}

D1, D2& WiCo-PG-base& 0.0238 &  &  \\ \cline{1-3}
D1, D3& WiCo-PG-base& 0.0220 &  &  \\ \cline{1-3}
D1 & WiCo-PG-base & 0.0345 &  &  \\ \hline

\end{tabular}
\label{scale}
\end{table}

\begin{table}[h!]
\centering
\caption{Ablation Study Results of WiCo-PG Under Different Module Configurations.}
\renewcommand{\arraystretch}{1.5}
\setlength{\tabcolsep}{6pt}
\small 
\begin{tabular}{|c|p{3.2cm}<{\centering}<{\centering}|}
\hline
\textbf{Configuration} & \textbf{NMSE on D1, D2, D3} \\ \hline
WiCo-PG-base & \textbf{0.0163}  \\ \hline
w/o frequency embedding & \underline{0.0222} \\ \hline
w/o S-R MoE & 0.0297  \\ \hline
\end{tabular}

\label{ablation}
\end{table}

\section{Conclusion}
In this paper, a novel WiCo for pathloss map generation, named WiCo-PG, has been developed to conduct massive and high-quality pathloss map generation for 6G AI-native communication system design. A new multi-modal sensing-communication dataset for WiCo-PG pre-training has been constructed, including 19.2\,k and 0.10\,B link-level pathloss values with diverse frequency bands and flight altitudes in multiple scenarios. To achieve massive and high-quality cross-modal pathloss generation, a dual-VQGAN and Transformer-based architecture has been proposed in WiCo-PG, and a two-stage pre-training strategy has been employed in WiCo-PG. Furthermore, to enhance generation and generalization performance, a frequency-guided S–R MoE Transformer architecture has been designed. Based on the designed architecture and pre-training strategy, the proposed WiCo-PG has shown improved pathloss map generation performance in multi-condition dataset unified learning, which has further improved with the expansion of data across diverse frequency bands and flight altitudes. Furthermore, validation across multiple scenarios has demonstrated that multi-condition unified learning has been essential to the performance enhancement of the proposed WiCo-PG. Simulation results have shown that the proposed WiCo-PG has possessed promising zero-shot generalization ability across unseen conditions. In the few-shot generalization, the proposed WiCo-PG has outperformed the LLM4PG and the GAN-based model by at least 1.37\,dB using no more than 2.7\% samples. Moreover, simulation results have revealed that, for cross-altitude and cross-band generalization, the WiCo-PG has generalized more effectively from low-altitude conditions with stronger blockage and low-frequency bands with more complex scattering to high-altitude and high-frequency conditions.



\begin{thebibliography}{1}
\bibliographystyle{IEEEtran}


\bibitem{channel modeling}
Y. Niu, Z. Wei, L. Wang, H. Wu, and Z. Feng, ``Interference management for integrated sensing and communication systems: A survey," \emph{IEEE Internet Things J.}, vol. 12, no. 7, pp. 8110--8134, Apr. 2025.


\bibitem{MMICM}
L. Bai, Z. Huang, M. Sun, X. Cheng, and L. Cui, ``Multi-modal intelligent channel modeling: A new modeling paradigm via Synesthesia of Machines," \emph{IEEE Commun. Surveys Tuts.}, early access, 2025.


\bibitem{large modeling}
R. Borralho, A. Mohamed, A. U. Quddus, P. Vieira, and R. Tafazolli, ``A survey on coverage enhancement in cellular networks: Challenges and solutions for future deployments," \emph{IEEE Commun. Surveys Tuts.}, vol. 23, no. 2, pp. 1302--1341, secondquarter 2021.

\bibitem{pathloss-twc}
B. Zhu, E. Bedeer, H. H. Nguyen, R. Barton, and Z. Gao, ``UAV trajectory planning for AoI-minimal data collection in UAV-aided IoT networks by Transformer," \emph{IEEE Trans. Wireless Commun.}, vol. 22, no. 2, pp. 1343--1358, Feb. 2023.

\bibitem{pathloss-twc2}
A. Hussain, A. Abdallah, and A. M. Eltawil, ``Redefining polar boundaries for near-field channel estimation for ultra-massive mimo antenna array," \emph{IEEE Trans. Wireless Commun.}, vol. 24, no. 10, pp. 8193--8207, Oct. 2025.


\bibitem{pathloss-twc3}
M. Mohammadi, L.-N. Tran, Z. Mobini, H. Q. Ngo, and M. Matthaiou, ``Cell-free massive MIMO-assisted SWIPT for IoT networks," \emph{IEEE Trans. Wireless Commun.}, vol. 24, no. 10, pp. 8598--8614, Oct. 2025.




\bibitem{scma1}
S. Ju, Y. Xing, O. Kanhere, and T. S. Rappaport, ``Millimeter wave and sub-terahertz spatial statistical channel model for an indoor office building," \emph{IEEE J. Sel. Areas Commun.}, vol. 39, no. 6, pp. 1561--1575, Jun. 2021.


\bibitem{cai1}
X. Cai \emph{et al.}, ``An empirical air-to-ground channel model based on passive measurements in LTE," \emph{IEEE Trans. Veh. Technol.}, vol. 68, no. 2, pp. 1140--1154, Feb. 2019.


\bibitem{scma2}
M. R. Akdeniz \emph{et al.}, ``Millimeter wave channel modeling and cellular capacity evaluation," \emph{IEEE J. Sel. Areas Commun.}, vol. 32, no. 6, pp. 1164--1179, Jun. 2014.



\bibitem{scma3}
W. Tang \emph{et al.}, ``Path loss modeling and measurements for reconfigurable intelligent surfaces in the millimeter-wave frequency band," \emph{IEEE Trans. Commun}, vol. 70, no. 9, pp. 6259--6276, Sept. 2022.






\bibitem{cai3}
X. Cai, X. Cheng, and F. Tufvesson, ``Toward 6G with terahertz communications: Understanding the propagation channels," \emph{IEEE Commun Mag.}, vol. 62, no. 2, pp. 32--38, Feb. 2024.



\bibitem{RT1}
K. Guan \emph{et al.}, ``Channel sounding and ray tracing for intrawagon scenario at mmWave and sub-mmWave bands,"  \emph{IEEE Trans. Antennas Propag.}, vol. 69, no. 2, pp. 1007--1019, Feb. 2021.

\bibitem{RT2}
K. Guan \emph{et al.}, ``Channel characterization for intra-wagon communication at 60 and 300 GHz bands," \emph{IEEE Trans. Veh. Technol.}, vol. 68, no. 6, pp. 5193--5207, Jun. 2019.


\bibitem{RF-driven1}
U. Masood, H. Farooq, A. Imran, and A. Abu-Dayya, ``Interpretable AI-based large-scale 3D pathloss prediction model for enabling emerging self-driving networks," \emph{IEEE Trans. Mobile Comput.}, vol. 22, no. 7, pp. 3967--3984, Jul. 2023.

\bibitem{RF-driven2}
Y. Yang \emph{et al.}, ``Generative-adversarial-network-based wireless channel modeling: Challenges and opportunities," \emph{IEEE Commun. Mag.}, vol. 57, no. 3, pp. 22--27, Mar. 2019.


\bibitem{AI-native}
F. Zhu \emph{et al.}, ``Wireless large AI model: Shaping the AI-native future of 6G and beyond,” arXiv preprint arXiv:2504.14653, 2025.


\bibitem{SoM}
X. Cheng \emph{et al.}, ``Intelligent multi-modal sensing-communication integration: Synesthesia of Machines," \emph{IEEE Commun. Surveys Tuts.}, vol.~26, no.~1, pp.~258--301, firstquarter 2024.


\bibitem{map1}
J.-H. Lee and A. F. Molisch, ``A scalable and generalizable pathloss map prediction," \emph{IEEE Trans. Wireless Commun.}, vol. 23, no. 11, pp. 17793--17806, Nov. 2024.

\bibitem{map2}
R. Levie, Ç. Yapar, G. Kutyniok, and G. Caire, ``RadioUNet: Fast radio map estimation with convolutional neural networks," \emph{IEEE Trans. Wireless Commun.}, vol. 20, no. 6, pp. 4001--4015, Jun. 2021.

\bibitem{map3}
C. Wang \emph{et al.}, ``Channel path loss prediction using satellite images: A deep learning approach,'' \emph{IEEE Trans. Mach. Learn. Commun. Netw.}, vol. 2, pp. 1357--1368, Sep. 2024.

\bibitem{U2G}
M. Sun, L. Bai, Z. Huang, and X. Cheng, ``Multi-modal sensing data based real-time path loss prediction for 6G UAV-to-ground communications,” \emph{IEEE Wireless Commun. Lett.}, vol. 13, no. 9, pp. 2462--2466, Sept. 2024.


\bibitem{uav-twc1}
X. Jing, F. Liu, C. Masouros, and Y. Zeng, "ISAC from the sky: UAV trajectory design for joint communication and target localization," \emph{IEEE Trans. Wireless Commun.}, vol. 23, no. 10, pp. 12857--12872, Oct. 2024.

\bibitem{uav-twc2}
N. Zhao, Z. Ye, Y. Pei, Y.-C. Liang, and D. Niyato, ``Multi-agent deep reinforcement learning for task offloading in UAV-assisted mobile edge computing," \emph{IEEE Trans. Wireless Commun.}, vol. 21, no. 9, pp. 6949--6960, Sept. 2022.


\bibitem{uav-twc3}
A. Khalili, A. Rezaei, D. Xu, F. Dressler, and R. Schober, ``Efficient UAV hovering, resource allocation, and trajectory design for ISAC with limited backhaul capacity," \emph{IEEE Trans. Wireless Commun.}, vol. 23, no. 11, pp. 17635--17650, Nov. 2024.





\bibitem{mr_gan}
M. Sun \emph{et al.}, ``A pathloss generation model for 6G dynamic U2G scenarios," \emph{IEEE Trans. Veh. Technol.}, submitted for publication, 2025.

\bibitem{small-no}
T. Jiao \emph{et al.}, ``Addressing the curse of scenario and task generalization in AI-6G: A multi-modal paradigm," \emph{IEEE Trans. Wireless Commun.}, vol. 24, no. 9, pp. 7377--7391, Sept. 2025.


\bibitem{LLM-yes}
Y. Chang \emph{et al.}, ``A survey on evaluation of large language models," \emph{ACM Trans. Intell. Syst. Technol.}, vol. 15, no. 3, pp. 1--45, Jun. 2024.

\bibitem{llm-twc}
X. Zhang \emph{et al.}, ``Beyond the cloud: Edge inference for generative large language models in wireless networks," \emph{IEEE Trans. Wireless Commun.}, vol. 24, no. 1, pp. 643--658, Jan. 2025.


\bibitem{llm-image}
J. Qin \emph{et al.}, ``DiffusionGPT: LLM-driven text-to-image generation system," \emph{arXiv preprint arXiv:2401.10061}, 2024.

\bibitem{llm-video}
X. Cao \emph{et al.}, ``Medical video generation for disease progression simulation," \emph{arXiv preprint arXiv:2411.11943}, 2024.


\bibitem{llm4wm}
X. Liu, S. Gao, B. Liu, X. Cheng, and L. Yang, ``LLM4WM: Adapting LLM for wireless multi-tasking," \emph{IEEE Trans. Mach. Learn. Commun. Netw.}, vol. 3, pp. 835--847, Jul. 2025.



\bibitem{llm-csi}
S. Fan \emph{et al.}, ``Csi-LLM: A novel downlink channel prediction method aligned with LLM pre-training,” \emph{arXiv preprint arXiv:2409.00005}, 2024.


\bibitem{llm-beam}
Y. Sheng \emph{et al.}, ``Beam prediction based on large language models," \emph{IEEE Wireless Commun. Lett.}, early access 2025.


\bibitem{LLM4PG}
M. Sun \emph{et al.}, ``LLM4PG: Adapting large language model for pathloss map generation via Synesthesia of
Machines," \emph{IEEE Trans. Mach. Learn. Commun. Netw.}, submitted for publication, 2025. 


\bibitem{channelgpt}
L. Yu \emph{et al.}, ``ChannelGPT: A large model toward real-world channel foundation model for 6G environment intelligence communication," \emph{IEEE Commun. Mag.}, vol. 63, no. 10, pp. 68--74, Oct. 2025.

\bibitem{tnse}
X. Cheng, B. Liu, X. Liu, E. Liu, and Z. Huang, ``Foundation model empowered Synesthesia of Machines (SoM): AI-native intelligent multi-modal sensing-communication integration,'' \emph{IEEE Trans. Network Sci. Eng.}, early access, 2025.


\bibitem{synthsom}
X. Cheng \emph{et al.}, ``SynthSoM: A synthetic intelligent multi-modal sensing-communication dataset for Synesthesia of Machines (SoM)”, \emph{Sci. Data}, vol. 12, pp. 819--833, May 2025.


\bibitem{AirSim}
S. Shah, D. Dey, C. Lovett, and A. Kapoor, ``AirSim: High-fidelity visual and physical simulation for autonomous vehicles,'' in \emph{Field and Service Robotics}, M. Hutter and R. Siegwart, Eds. Cham, Switzerland: Springer, 2018, pp. 621–635.

\bibitem{Wireless InSite}
\emph{Remcom}. Wireless InSite. [Online]. Available: https://www.remcom.com/wireless-insite-em-propagation-software [Publication date: Jan. 2017, Accessed date: Mar. 2022].






\bibitem{vqgan}
P. Esser, R. Rombach, and B. Ommer, ``Taming transformers for high-resolution image synthesis,” in \emph{Proc. IEEE CVPR'21}, Nashville, TN, USA, Jun. 2021, pp. 12873--12883.

\bibitem{pl}
J. Bian, C.-X. Wang, X. Gao, X. You, and M. Zhang, ``A general 3D non-stationary wireless channel model for 5G and beyond," \emph{IEEE Trans. Wireless Commun.}, vol. 20, no. 5, pp. 3211--3224, May 2021.



\bibitem{adam}
D. P. Kingma and J. Ba, ``Adam: A method for stochastic optimization,” in \emph{Proc. ICLR'15}, May 2015, pp. 1–12.

\bibitem{blockage-twc}
J. Wang \emph{et al.}, ``Covert air-ground relaying with blockages," \emph{IEEE Trans. Wireless Commun.}, vol. 24, no. 10, pp. 8347--8360, Oct. 2025.




\end{thebibliography}
%

\newpage

 





\end{document}